\documentclass[11pt]{article}

\usepackage[margin=1in]{geometry}
\usepackage[T1]{fontenc}
\usepackage{amsmath,amssymb}
\usepackage{graphicx}
\usepackage{booktabs}
\usepackage{xcolor}
\usepackage{authblk}
\usepackage{cite}
\usepackage{hyperref}

\newif\ifrevisedon\revisedontrue
\makeatletter
\long\def\revised#1{\ifrevisedon{\color{black}#1}\else#1\fi}
\makeatother

\hypersetup{%
	pdfauthor={Shrenik Jadhav, Vidhyashree Nagaraju, Zheng Liu},
	pdftitle={Data Center Life Cycle Co-Design Optimization},
	pdfkeywords={Data center thermal management, cooling subloop architecture, co-design, life cycle assessment, multi-objective optimization, reliability},
	pdfsubject = {Cooling plant subloop architecture co-design optimization},
	colorlinks=true,
	linkcolor=blue,
	citecolor=blue,
	urlcolor=blue,
}

\title{Data Center Life Cycle Co-Design Optimization}

\author[1]{Shrenik Jadhav}
\author[2]{Vidhyashree Nagaraju}
\author[3]{Zheng Liu\thanks{Corresponding author. Email: \texttt{zhengtl@umich.edu}}}

\affil[1]{Department of Computer and Information Science, University of Michigan-Dearborn, 4901 Evergreen Rd, Dearborn, MI, USA. Current address: Department of Electrical Engineering and Computer Science, Embry-Riddle Aeronautical University, 1 Aerospace Blvd, Daytona Beach, FL, USA}
\affil[2]{Department of Electrical Engineering and Computer Science, Embry-Riddle Aeronautical University, 1 Aerospace Blvd, Daytona Beach, FL, USA}
\affil[3]{ASME member. Department of Industrial and Manufacturing Systems Engineering, University of Michigan-Dearborn, 4901 Evergreen Rd, Dearborn, MI, USA}

\date{}

\begin{document}

\maketitle

\begin{abstract}
Liquid cooled supercomputers dissipate tens of megawatts of waste heat through cooling plants organized as parallel subloops that serve coolant distribution units. The number of subloops and the assignment of units to them are design decisions fixed at construction, yet they have not been systematically optimized at this scale. \revised{We present a framework that integrates operational energy from a validated control optimizer, embodied carbon and capital cost from a bill of materials, maintenance over the service life, and expected unplanned downtime from a component level reliability model. All 611 ways of partitioning the 25 coolant distribution units of the Frontier supercomputer into two through six subloops are evaluated. When redundancy is not costed, the optimum is two subloops holding 14 and 11 units, at 3,320.7 tonnes of carbon dioxide equivalent and \$3,987k over a 7 year horizon, saving 35.7 tonnes and \$63k against the documented as built configuration of three duty subloops holding 14, 6 and 5 units. The difference is driven by piping rather than by operational energy, and the identity of the optimum is unchanged across 15 sensitivity scenarios and both extremes of physical unit grouping, although its margin narrows under compact grouping. When the N+1 standby train that the plant actually carries is priced, the optimum moves to four or five duty subloops, adjacent to the three the plant runs and far from the unconstrained answer, and the semi-analytical decision rule reproduces this shift across four leadership class systems. Redundancy policy, not cost or carbon, is what sets the subloop count. A conversion of the built plant is shown not to pay back, so the framework is a greenfield design tool.}
\end{abstract}

\noindent\textbf{Keywords:} Data center thermal management, \revised{cooling subloop architecture}, co-design, life cycle assessment, multi-objective optimization, reliability

\section{Introduction}
\label{sec:introduction}

Data center electricity consumption has grown sharply over the past decade and is forecast to grow further. The United States Department of Energy estimates that domestic data centers used 4.4\% of total US electricity in 2023 and projects this share to rise to between 6.7 and 12\% by 2028, driven primarily by artificial intelligence (AI) workloads \cite{shehabi2024lbnl}. The International Energy Agency projects that global data center electricity demand could exceed 1,000 TWh by 2026, roughly double its 2022 level \cite{iea2024electricity}. Cooling systems are a large share of this consumption. For typical air and water cooled facilities, cooling accounts for 33 to 40\% of total data center energy use \cite{sun2024frontier}. For exascale supercomputers such as Frontier at Oak Ridge National Laboratory (ORNL), peak power demand reaches 30 MW with a corresponding cooling load that must be removed continuously through liquid cooling infrastructure \cite{sun2024frontier}.

As the electricity grid decarbonizes, the relative importance of embodied carbon in data center facilities rises. \revised{Embodied carbon is the greenhouse gas released in extracting, processing, transporting and installing the materials and equipment that make up a facility, counted once at construction and reported here in tonnes of carbon dioxide equivalent (t CO\textsubscript{2}e). Facility carbon is the sum of that embodied term and the operational carbon released by the electricity the facility draws over its service life.} Construction materials, mechanical equipment, and balance of plant components are responsible for an increasing share of life cycle greenhouse gas emissions, with some recent estimates placing embodied carbon at 38 to 69\% of total facility carbon over a 60 year horizon \cite{imasons2024lca}. Design decisions made before construction thus have a larger and more persistent climate impact than they did under older grid carbon assumptions. \revised{This shift is the reason design stage methods that carry life cycle consequences into the decision itself have become central to design research \cite{ramani2010,kota2010}, and the reason of allocation and end of life conventions have to be stated rather than assumed \cite{bouchouireb2021}.} Cooling plant topology is one such design decision: once a facility is built with a fixed number of cooling loops, pumps, and headers, retrofitting the topology is expensive and disruptive.

\revised{Research on data center cooling spans four levels, and the design questions differ at each. At the chip and server level the concern is heat removal from a package: direct to chip cold plates have been characterized experimentally at high heat density and shown to hold junction temperatures at coolant conditions that air cooling cannot reach \cite{heydari2024coldplate}. At the rack and row level the concern is the interface between the information technology equipment and the facility: rear door heat exchangers have been analyzed for the trade between thermal performance and fan energy \cite{cho2024rdhx}, and liquid to liquid coolant distribution units (CDUs), the components this study partitions, have been characterized for capacity, approach temperature and pumping penalty in an in-row configuration \cite{shahi2025cdu}. At the room level, airflow management, containment and provisioning have been reviewed extensively for air cooled halls \cite{chu2019airflow}. At the facility level, the plant itself is the object of design: surveys of cooling technology, power consumption modeling and control strategy set out the available architectures and the models used to size and operate them \cite{zhang2021survey}, open source Modelica libraries support simulation based comparison of chiller plant and water side economizer control \cite{fan2021modelica}, parallel pumping systems have been studied for the control strategy that minimizes energy at a given duty \cite{viholainen2012pumps}, and waste heat from the plant has been evaluated as a district heating resource \cite{wahlroos2017wasteheat}.}

\revised{Two threads in this literature are close to the present problem without addressing it. The first is plant topology optimization in adjacent domains: district cooling network design has been posed as a mixed integer program over which loads connect to which network branch, with heuristics used where the exact program does not scale \cite{neri2022district}. That work optimizes network connection for a distribution network at city scale; the analogous question inside a single exascale cooling plant, how many parallel subloops to build and which CDUs to place on each, has not been posed. The second is design methodology: life cycle objectives have been carried into discrete design and allocation problems, for instance in jointly optimizing vehicle design and its deployment to minimize life cycle greenhouse gas emissions \cite{shiau2011}, reliability has been treated as a system level constraint on a design optimization rather than as a separate verification step \cite{mcdonald2008}, Pareto set structure has been studied as a design decision aid \cite{unal2018}, energy optimization has been posed over building clusters where supply and demand are coupled \cite{odonkor2016}, and surrogate assisted search has been developed for cases where each design evaluation is expensive \cite{bhattacharjee2018}. The present study applies this methodological apparatus to a plant design problem that has so far been settled by convention.}

Work specific to the Frontier cooling plant has progressively sharpened the operational picture. A physics guided machine learning framework with a monotonicity constrained gradient boosting surrogate predicts facility accessory power and, used as a baseline, quantifies about 85 MWh of annual cooling inefficiency \cite{Jadhav2026MLGuided}. A Modelica digital twin of the Frontier high temperature water (HTW) cooling system, validated through one full calendar year of 10 min operational data per ASHRAE Guideline 14 \cite{ashrae2014g14}, supports operational strategies that reach up to 35.5\% total cooling energy savings \cite{Jadhav2026Stage1}. A further study develops joint flow fraction and supply temperature co-design optimization across all CDU partitions, focusing on the operational side of the design space \cite{Jadhav2026CoDesign}. \revised{System scale power and thermal behavior of leadership computing facilities has also been characterized directly from instrumentation \cite{shin2021,karimi2024}, which is the measurement basis this line of work depends on.}

What remains unaddressed is the upstream design choice of how many subloops to build and how to assign CDUs to them. The number of parallel subloops, the assignment of CDUs to subloops, and the resulting equipment and piping requirements together determine the embodied carbon and capital cost of the plant. Operational savings from improved control are bounded by what the chosen topology allows. Yet no published study enumerates the space of feasible cooling subloop configurations for an exascale system, evaluates each on both operational and embodied carbon, and identifies the optimum jointly.

\revised{This paper closes that gap. We build a three layer framework that enumerates every feasible subloop topology for a 25 CDU exascale plant, evaluates each on operational energy from a validated per timestep optimizer, on embodied carbon and capital cost from a bill of materials, on maintenance labor and replacement parts over the service life, and on expected unplanned downtime from a component level reliability model, and then aggregates these into a single life cycle objective. Applied to Frontier, the framework locates the cost and carbon optimum, decomposes what drives it, and shows that the driver is piping mass rather than operational energy. It then shows that this ranking reverses once the standby train the plant actually carries is priced, which moves the recommended duty count from two, far below what the plant runs, to four or five, just above it. A semi-analytical rule reduces the whole calculation to two system parameters and a stated redundancy policy, and reproduces the shift for three further leadership class systems. Finally the framework is tested where it would be used: we measure how much operational data the design decision needs, how the search behaves as plants grow, how the result moves when the physical grouping of CDUs is varied between its extremes, and whether converting an existing plant would pay. Section \ref{sec:problem} introduces the systems, the decision and the notation; Section \ref{sec:methods} develops the framework; Section \ref{sec:results} presents the results; Section \ref{sec:discussion} draws the design implications and the limits; Section \ref{sec:conclusions} concludes.}

\begin{figure}[t]
\centering
\includegraphics[width=\textwidth]{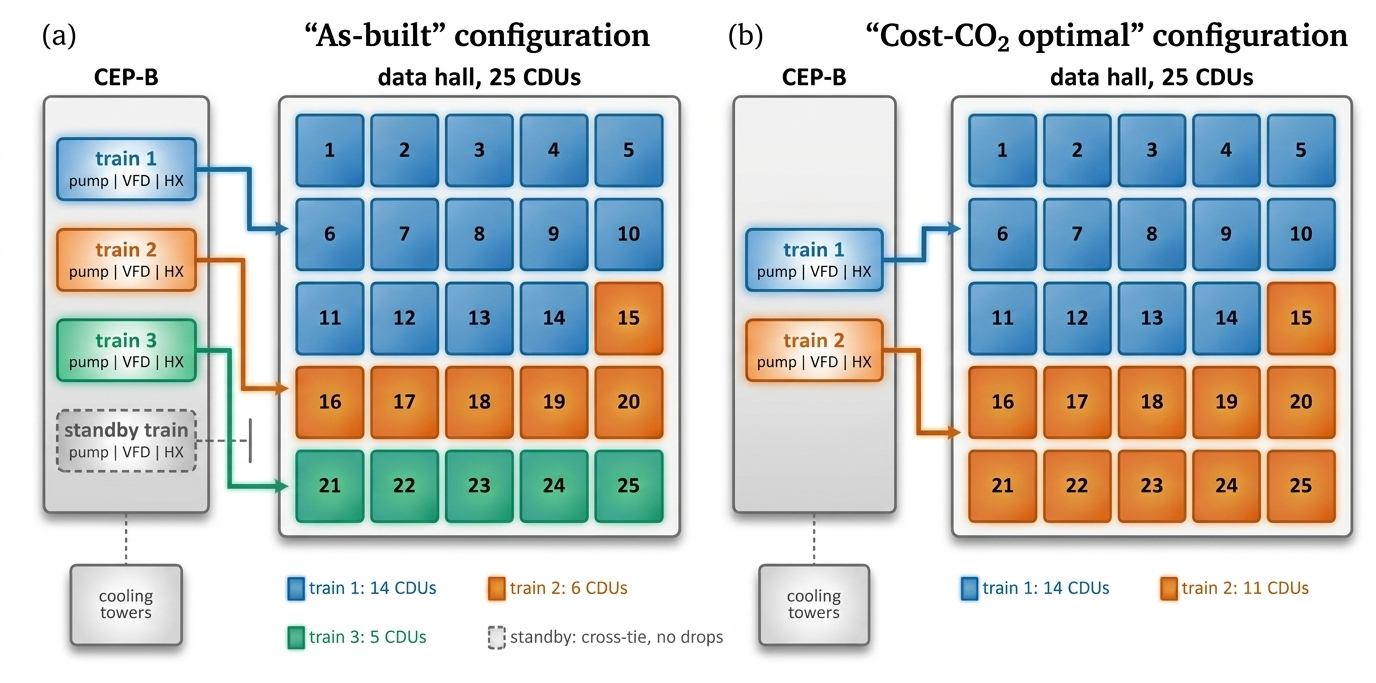}
\caption{\revised{Frontier HTW subloop topology, redrawn for this revision. (a) As built: three duty trains carrying the documented in-service partition $(14,6,5)$, with a fourth train held as a standby that is cross-tied to the header and carries no dedicated drops. (b) Life cycle cost and carbon optimum, the two duty trains of $(14,11)$; a standby would be added to this design under the N+1 policy. Each train serves a contiguous block of CDUs, and its connector terminates at the vertical center of the block it serves.}}
\label{fig:topology}
\end{figure}

\revised{
\section{Problem Overview}
\label{sec:problem}

\subsection{Systems Considered}
\label{sec:systems}

Four liquid cooled high performance computing (HPC) systems appear throughout this paper. They are not interchangeable, and the differences matter for the design question, so they are set out here rather than assumed. Table \ref{tab:systems} lists them.

\begin{table}[t]
\caption{\revised{Systems considered. All four are 100\% direct liquid cooled. Performance and power are TOP500 reported values \cite{top500}; cooling parameters marked with an asterisk are inferred from public architectural descriptions \cite{olcf_frontier} and are treated as estimates throughout.}}
\label{tab:systems}
\centering
\small
\begin{tabular}{lcccccccc}
\toprule
System & Site & First & Class & $R_{\mathrm{max}}$ & Power & HTW peak & CDUs & Duty subloops \\
 &  & operation &  & (PFlop/s) & (MW) & flow (kg/s) & $N$ & as built \\
\midrule
Frontier   & OLCF & 2022 & exascale     & 1,353 & 24.6 & 376 & 25 & 3 duty, 1 standby \\
Aurora     & ALCF & 2023 & exascale     & 1,012 & 38.7 & 756\textsuperscript{*} & 42\textsuperscript{*} & not disclosed \\
El Capitan & LLNL & 2024 & exascale     & 1,742 & 30.0 & 378\textsuperscript{*} & 22\textsuperscript{*} & not disclosed \\
LUMI       & CSC  & 2022 & pre-exascale &   379 &  7.1 &  76\textsuperscript{*} & 10\textsuperscript{*} & not disclosed \\
\bottomrule
\end{tabular}
\end{table}

All four are direct liquid cooled machines built on the same broad plant pattern, so the design question posed here applies to each. They differ by roughly an order of magnitude in peak flow, which is the parameter the decision turns on. Only Frontier has a published operational record at the resolution this study requires \cite{sun2024frontier}, so it is the case study and the other three are used to test whether the resulting design rule transfers.

\subsection{The Cooling Loops and the Decision}
\label{sec:loops}

Heat leaves an exascale machine through a chain of loops. Inside each compute blade, coolant passes over cold plates on the processors and memory. That coolant circulates in the technology cooling system, a closed loop served by a CDU. Each Frontier CDU cools three rack cabinets, and there are $N=25$ of them in the data hall \cite{sun2024frontier}. On the facility side of each CDU is the HTW loop, which carries heat from the data hall to the central energy plant designated CEP-B, where plate heat exchangers pass it to the condenser water loop and mechanical draft cooling towers reject it to atmosphere.

The HTW loop is not a single circuit. It is divided into $K$ parallel subloops, each with its own pump, variable frequency drive (VFD), plate heat exchanger and isolation valve set, and each serving a subset of the 25 CDUs through a header and a set of drops. Frontier's CEP-B contains four such subloops, of which three are in service and the fourth is held as a backup \cite{sun2024frontier}; the three duty subloops serve 14, 6 and 5 CDUs respectively \cite{Jadhav2026CoDesign}.

Three decisions define this architecture, and they are made at different times. The subloop count $K$ and the assignment of CDUs to subloops are fixed at construction and are the design variables of this paper. The flow fraction sent to each subloop and the HTW supply temperature setpoint are control variables, adjusted continuously in operation. The redundancy policy, meaning whether a standby train is provided and what it must be able to take over, is a facility policy decision that constrains the design variables. Figure \ref{fig:topology} shows two candidate architectures for the same 25 CDUs.

A count based partition specifies how many CDUs each subloop serves, not which physical CDUs are grouped. Since piping length and hydraulic loss depend on the grouping, each partition maps to a family of physical layouts. Section \ref{sec:layout} brackets that family explicitly.

\subsection{Notation}
\label{sec:notation}

The integer partition $\mathbf{n} = (n_1, \ldots, n_K)$ assigns $n_k$ CDUs to subloop $k$, with $\sum_{k=1}^K n_k = N = 25$. The flow fraction $f_k$ is the fraction of total HTW flow directed to subloop $k$, with $\sum_{k=1}^K f_k = 1$. The per CDU heat load at timestep $t$ is $w_i^{(t)}$, and the subloop heat load is $W_k^{(t)} = \sum_{i \in S_k} w_i^{(t)}$ where $S_k$ is the set of CDUs on subloop $k$. Total HTW mass flow is $\dot{m}$ and supply temperature is $T_s$. The partition space has $|\mathcal{P}_2| = 12$, $|\mathcal{P}_3| = 52$, $|\mathcal{P}_4| = 120$, $|\mathcal{P}_5| = 192$, $|\mathcal{P}_6| = 235$ elements, 611 in total.
}

\begin{figure}[t]
\centering
\includegraphics[width=0.98\textwidth]{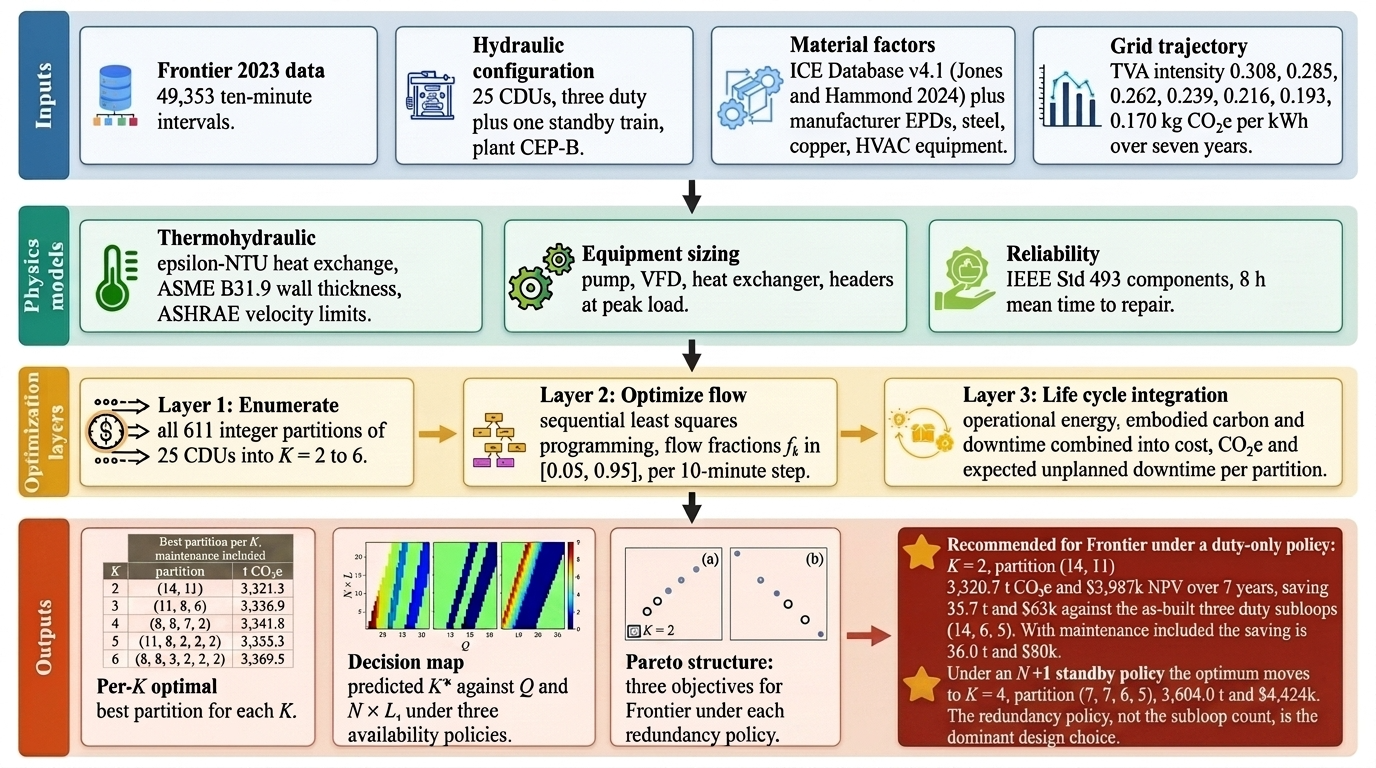}
\caption{\revised{Three layer optimization framework. Layer 1 enumerates every feasible subloop topology, Layer 2 optimizes plant operation for each, and Layer 3 aggregates energy, carbon, cost and downtime into one life cycle objective per design. The availability policy enters as an input and is revised as a scenario rather than being fixed.}}
\label{fig:framework}
\end{figure}

\section{Methodology}
\label{sec:methods}

\revised{This section develops the framework. Section \ref{sec:framework_overview} gives the top down view, the system boundary and the functional unit. Sections \ref{sec:plant_model} through \ref{sec:reliability_model} give the four submodels: cooling plant and operational energy, embodied carbon and capital cost, maintenance, and reliability. Section \ref{sec:multi_obj} defines the life cycle aggregation and the Pareto formulation, Section \ref{sec:decision_rule} gives the semi-analytical decision rule and how the redundancy policy enters it, and Section \ref{sec:sensitivity_design} describes the sensitivity design. The notation used throughout is in Section \ref{sec:notation}.}

\subsection{Framework Overview, System Boundary and Functional Unit}
\label{sec:framework_overview}

\revised{
\begin{figure}[t]
\centering
\includegraphics[width=0.95\textwidth]{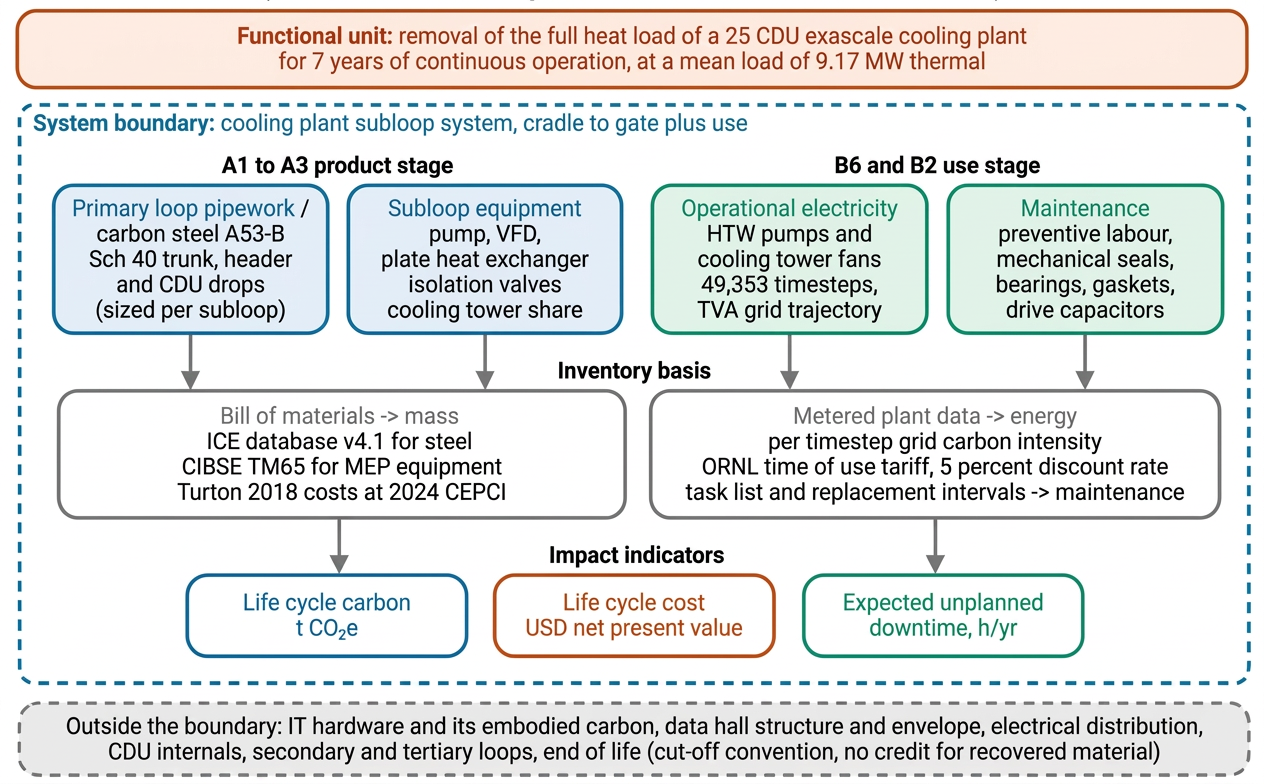}
\caption{\revised{Life cycle system boundary and functional unit. The assessment covers the subloop system only: primary loop pipework, per subloop equipment, the operational electricity that drives it, and the maintenance it requires. Information technology hardware, the data hall structure, electrical distribution, CDU internals and the secondary and tertiary loops are outside the boundary because they do not change with the subloop count.}}
\label{fig:boundary}
\end{figure}

The functional unit is the removal of the full heat load of a 25 CDU exascale cooling plant for 7 years of continuous operation, at the mean thermal load of 9.17 MW measured in the 2023 record. Every candidate topology is required to deliver this same function, so the comparison is like for like. The 7 year horizon matches HPC system replacement cadence rather than building services practice, and is swept at 5 and 10 years.

Figure \ref{fig:boundary} shows the system boundary. It follows the modular structure of EN 15978 \cite{en15978} and the general framework of ISO 14040 and 14044 \cite{iso14040,iso14044}. Product stage modules A1 to A3 cover the primary loop pipework and the per subloop equipment. Use stage module B6 covers operational electricity and module B2 covers maintenance. End of life is treated under the cut-off convention: material leaving the boundary carries no credit. This choice is stated because it matters in one place, the retrofit assessment in Section \ref{sec:retrofit}, where crediting recovered steel would reverse the sign of the carbon result; both conventions are reported there.

Everything inside the boundary is a quantity that changes when the subloop count or the CDU assignment changes. Information technology hardware and its embodied carbon, the data hall structure and envelope, electrical distribution, the CDU internals, and the secondary and tertiary loops are outside the boundary because they are identical across candidates and would cancel from every comparison.
}

Figure \ref{fig:framework} summarizes the framework. The inputs are the Frontier operational record from 2023 covering 49,353 ten-minute intervals \cite{sun2024frontier}, the plant configuration, the embodied carbon factors for steel, copper, and heating, ventilation, and air conditioning (HVAC) equipment from the Inventory of Carbon and Energy (ICE) database \cite{hammond2008ice,ice_v41} supplemented with manufacturer environmental product declarations (EPDs) \cite{iso14025} and the mechanical plant methodology of CIBSE TM65 \cite{cibse_tm65}, the Tennessee Valley Authority (TVA) grid intensity trajectory declining from 0.308 kg CO\textsubscript{2}e per kWh in 2024 to 0.17 in 2030 and 0.11 in 2035 with a floor near 0.08 thereafter, \revised{and the availability policy}.

Four physics models translate inputs into per-partition quantities: a thermohydraulic model using the effectiveness-number of transfer units ($\varepsilon$-NTU) method, an equipment and pipe sizing model, \revised{a maintenance model, and a reliability model}. Three optimization layers then act on them. Layer 1 enumerates all 611 integer partitions of $N=25$ into $K \in \{2,3,4,5,6\}$ subloops. Layer 2 invokes the sequential least squares programming (SLSQP) optimizer of \cite{Jadhav2026CoDesign} to compute the optimal flow allocation and supply temperature at each 10 min timestep. \revised{Layer 3 aggregates energy, carbon, cost and downtime into a single life cycle objective for every candidate.}

\subsection{Cooling Plant Model and Operational Energy}
\label{sec:plant_model}

\revised{This section explains the plant model and the operational energy sub-model together.} The framework inherits the reduced order plant model validated in our previous work \cite{Jadhav2026Stage1} through one calendar year of operational data, achieving a coefficient of variation of the root mean square error (CV-RMSE) within 2.67\% and a normalized mean bias error (NMBE) within $\pm 2.5$\% per ASHRAE Guideline 14 \cite{ashrae2014g14}.

\revised{Three relations define the model. Total HTW pump power follows the affinity law for variable speed centrifugal pumps operating against a quadratic system curve.

\begin{equation}
P_{\mathrm{pump}}(\dot{m}) = P_{\mathrm{pump,nom}} \left( \frac{\dot{m}}{\dot{m}_{\mathrm{nom}}} \right)^{n_p}
\label{eq:pump_power}
\end{equation}
where $\dot{m}_{\mathrm{nom}} = 190$ kg/s, $P_{\mathrm{pump,nom}} = 17.18$ kW and $n_p = 3$, all calibrated to the digital twin. Cooling tower fan power follows an empirical correlation in heat rejection and approach temperature, Eq. \eqref{eq:fan_power}.

\begin{equation}
P_{\mathrm{fan}}(\dot{m}, T_s) = P_{\mathrm{fan,nom}} \, \frac{Q_{\mathrm{rej}}}{Q_{\mathrm{rej,nom}}} \cdot \frac{\Delta T_{\mathrm{app,nom}}}{\Delta T_{\mathrm{app,base}} + (T_s - T_{s,\mathrm{base}})}
\label{eq:fan_power}
\end{equation}
where $P_{\mathrm{fan,nom}} = 950.13$ kW, $Q_{\mathrm{rej,nom}} = 9{,}170.7$ kW, $\Delta T_{\mathrm{app,nom}} = 4.0$ K and a calibrated baseline offset $\Delta T_{\mathrm{app,base}} = 23.5$ K, all inherited from the calibration of the companion study \cite{Jadhav2026CoDesign}. Equation \eqref{eq:fan_power} is an empirical correlation calibrated at the measured operating point rather than a first principles fan affinity model, and its validity range is the supply temperature envelope the optimized strategies occupy. Raising the supply temperature widens the approach and reduces fan power, which is the mechanism the co-design strategies exploit. The per subloop return temperature follows from a lumped energy balance is shown in Eq \eqref{eq:return_temp}.

\begin{equation}
T_{\mathrm{ret},k}^{(t)} = T_s^{(t)} + \frac{W_k^{(t)}}{f_k^{(t)} \, \dot{m}^{(t)} \, c_p}
\label{eq:return_temp}
\end{equation}
where $c_p = 3{,}500$ J/(kg K) for the working fluid.}

\revised{At each timestep the optimizer minimizes total cooling electrical power over the flow fractions, the total flow and the supply temperature.

\begin{equation}
\begin{aligned}
\min_{\mathbf{f},\,\dot m,\,T_s}\quad & P_{\mathrm{pump}}(\dot m) + P_{\mathrm{fan}}(\dot m, T_s) \\
\text{s.t.}\quad & T_{\mathrm{ret},k} \le T_{\mathrm{lim}} = 42\,^{\circ}\mathrm{C}, \quad k = 1,\ldots,K, \\
& \textstyle\sum_{k} f_k = 1, \quad f_k \in [0.05,\,0.95], \\
& \dot m \in [30,\,420]\ \mathrm{kg/s}, \quad T_s \in [10,\,35]\,^{\circ}\mathrm{C}.
\end{aligned}
\label{eq:operational}
\end{equation}

The objective is the total electrical power drawn by the cooling plant, which is what the facility pays for and what the grid emits against. The two terms pull in opposite directions and that opposition is the whole content of the problem: raising $T_s$ widens the cooling tower approach and cuts fan power, which dominates the plant at about 73\% of cooling electricity, but it also raises every return temperature through Eq. \eqref{eq:return_temp} and so pushes the constraint. Recovering the margin requires more flow, which costs pump power cubically through Eq. \eqref{eq:pump_power}. The optimizer sits at the point where the marginal fan saving from one more degree of supply temperature equals the marginal pumping cost of the flow needed to absorb it, subject to the hottest subloop staying below 42 $^{\circ}$C. The flow fractions matter because that constraint binds on one subloop at a time: a subloop carrying a disproportionate share of the heat load binds first and forces the whole plant to a lower supply temperature, and redistributing flow towards it releases the plant. This is the mechanism through which the partition enters the operational term at all.}

\revised{The limits are physical. The return limit $T_{\mathrm{lim}}$ is the equipment protection limit of 45 $^{\circ}$C less a 3 $^{\circ}$C commissioning margin. The lower flow fraction bound prevents a subloop from being starved to the point where the heat exchanger correlation loses validity; it is applied per subloop, so higher $K$ reserves more total minimum flow, which if anything penalizes higher $K$ and means the low $K$ result is not an artifact of the bound. The flow and temperature ranges are the operating envelope of the plant as commissioned.}

Three operating strategies are distinguished, following \cite{Jadhav2026CoDesign}. Strategy A optimizes flow only. Strategy B co-optimizes flow and supply temperature with flow fractions held proportional to CDU count. Strategy C adds per timestep optimization of the flow fractions. Strategy C is the production configuration and is used throughout unless stated. Annual energy is the integral of Eq. \eqref{eq:operational} over the 49,353 timesteps. Annual operational carbon multiplies per timestep demand by the per timestep grid intensity; annual operational cost multiplies it by the ORNL time of use tariff. Life cycle totals use a 7 year horizon with future cash flows discounted at 5\% annually.

\subsection{Embodied Carbon and Capital Cost Model}
\label{sec:embodied_model}

The volumetric flow carried by subloop $k$ at peak is shown in Eq. \eqref{eq:subloop_flow}.

\begin{equation}
Q_k = \frac{n_k}{N}\,\frac{\dot m_{\mathrm{peak}}}{\rho_w}
\label{eq:subloop_flow}
\end{equation}
where $\dot m_{\mathrm{peak}} = 376$ kg/s and $\rho_w = 994$ kg/m\textsuperscript{3}. From Eq. \eqref{eq:subloop_flow}, the nominal bore is the smallest Schedule 40 size in ASME B36.10M \cite{asme_b3610m} whose inside diameter satisfies Eq. \eqref{eq:bore_select}.

\begin{equation}
\frac{\pi D_{i}^{2}}{4} \ \ge\ \frac{Q_k}{v_{\mathrm{des}}}
\label{eq:bore_select}
\end{equation}
where $v_{\mathrm{des}}$ is the hydronic design velocity for that size class from ASHRAE Fundamentals \cite{ashrae_fundamentals}, 0.91, 1.83 or 2.44 m/s for small, medium and large bore. The selected wall thickness is checked against the minimum of ASME B31.9 \cite{asme_b31_9}, Eq. \eqref{eq:b319}.

\begin{equation}
t_{\min} = \frac{P_d D_{o}}{2\,(S E + P_d Y)} + A
\label{eq:b319}
\end{equation}
where design pressure $P_d = 1$ MPa, allowable stress $S = 103.4$ MPa for A53 Grade B, joint factor $E = 1$, coefficient $Y = 0.4$ and corrosion allowance $A = 1.5$ mm, divided by the 0.875 mill tolerance. Every selection in the enumeration passes this check with margin, because the velocity criterion governs at these pressures.

Mass per unit length is shown in Eq. \eqref{eq:mass_per_m}.
\begin{equation}
\mu_k = \frac{\pi}{4}\left(D_{o}^{2} - D_{i}^{2}\right)\rho_s
\label{eq:mass_per_m}
\end{equation}
where $\rho_s = 7{,}850$ kg/m\textsuperscript{3}. The routed length of subloop $k$, counting supply and return.

\begin{equation}
L_k = 2 L_{\mathrm{trunk}} + 2 L_{\mathrm{header}} + 2 L_{\mathrm{drop}} n_k
\label{eq:route_len}
\end{equation}
where $L_{\mathrm{trunk}} = 100$ m, $L_{\mathrm{header}} = 50$ m and $L_{\mathrm{drop}} = 5$ m. Total pipe mass is $M_{\mathrm{pipe}} = \sum_k \mu_k L_k$, from Eqs. \eqref{eq:mass_per_m} and \eqref{eq:route_len}. Because each subloop replicates its own trunk and header, and because bore is quantized to standard schedules, $M_{\mathrm{pipe}}$ rises with $K$ even though the flow per subloop falls. Section \ref{sec:layout} replaces the fixed header allowance with an explicit floor plan and reports the effect.

Pump shaft power and heat exchanger area for subloop $k$ are given by Eq. \eqref{eq:equip},
\begin{equation}
P_{s,k} = \frac{\dot m_k\, g\, H}{\eta_p}, \qquad
A_k = \frac{\dot m_k\, c_p\, \Delta T_{\mathrm{peak}}}{U\,\Delta T_{\mathrm{lm}}}
\label{eq:equip}
\end{equation}
where total dynamic head $H = 120$ m, pump efficiency $\eta_p = 0.78$, motor efficiency 0.95, peak temperature rise $\Delta T_{\mathrm{peak}} = 19$ K, overall coefficient $U = 4{,}500$ W/(m\textsuperscript{2} K) and log mean difference $\Delta T_{\mathrm{lm}} = 5$ K. The VFD is rated at 1.1 times motor input power, and the cooling tower share is allocated in proportion to CDU count.

Embodied carbon and capital cost are
\begin{align}
G_{\mathrm{emb}} &= e_{\mathrm{pipe}} M_{\mathrm{pipe}} + \sum_k \bigl[ e_{\mathrm{pump}} P_{s,k} \notag\\
 &\quad + e_{\mathrm{vfd}} P_{v,k} + e_{\mathrm{hx}} A_k + e_{\mathrm{ct}} \dot Q_k \bigr], \label{eq:emb_co2}\\
C_{\mathrm{cap}} &= c_{\mathrm{pipe}} F_{\mathrm{BM}} M_{\mathrm{pipe}} + \sum_k \bigl[ c_{\mathrm{pump}}(P_{s,k}) \notag\\
 &\quad + c_{\mathrm{vfd}} P_{v,k} + c_{\mathrm{hx}} A_k + c_{\mathrm{ct}} \dot Q_k \bigr] \label{eq:emb_cost}
\end{align}

In Eqs. \eqref{eq:emb_co2} and \eqref{eq:emb_cost}, with $e_{\mathrm{pipe}} = 3.0$ kg CO\textsubscript{2}e/kg from the ICE database \cite{ice_v41}, and factors derived from CIBSE TM65 \cite{cibse_tm65} of $e_{\mathrm{pump}} = 100$ kg CO\textsubscript{2}e/kW, $e_{\mathrm{vfd}} = 1$ kg CO\textsubscript{2}e/kW, $e_{\mathrm{hx}} = 450$ kg CO\textsubscript{2}e/m\textsuperscript{2} and $e_{\mathrm{ct}} = 12$ kg CO\textsubscript{2}e/kW thermal. Pipe installation uses $F_{\mathrm{BM}} = 3.5$, and pump cost follows the Turton correlation \cite{turton2018} escalated from CEPCI 567.5 to 800.

\revised{Worked numerically for the $K=2$ partition $(14,11)$: subloop 1 carries $Q_1 = 0.212$ m\textsuperscript{3}/s, selecting NPS 14 Schedule 40 at $\mu_1 = 94.3$ kg/m over its routed length $L_1 = 440$ m for 41,493 kg; subloop 2 carries $Q_2 = 0.166$ m\textsuperscript{3}/s, selecting NPS 12 at $\mu_2 = 79.7$ kg/m over $L_2 = 410$ m for 32,693 kg. A single bore is carried through the trunk, header and drops of each subloop. Total pipe mass is 74,186 kg, which at the central factor of 3.0 kg CO\textsubscript{2}e/kg gives the pipe term of 222.6 t; the equipment terms add 1,134.4 t, for 1,356.9 t of embodied carbon.}

\revised{A standby train, where the redundancy policy of Section \ref{sec:decision_rule} calls for one, is costed on the same basis with one exception. It carries a trunk and a header sized to take over the largest duty branch, but no dedicated CDU drops, because it serves no units of its own in normal operation and is cross-tied to the headers of the duty trains. Its embodied carbon and capital cost enter through Eqs. \eqref{eq:emb_co2} and \eqref{eq:emb_cost} as an additional term whose size is set by the largest $n_k$ in the partition.}

Because each subloop is sized to the flow it carries, and the per subloop flows sum to the fixed system total, total equipment embodied carbon is invariant to $K$ at 1,134 t for every partition. The only embodied term that changes with $K$ is piping.

\revised{
\subsection{Maintenance Model}
\label{sec:maintenance}

Maintenance scales with the number of maintainable trains rather than with plant flow, so it is a function of $K$ and belongs in the life cycle objective.

Preventive maintenance per train per year comprises quarterly pump inspection, an annual alignment and vibration survey, an annual drive thermal scan and filter change, an annual heat exchanger differential pressure and gasket check, and an annual valve stroke and packing check. The task list totals 17 h per train per year plus \$350 of consumables. Scheduled replacement covers the mechanical seal at a 3 year interval, pump bearings and the heat exchanger gasket set at 5 years, and the drive capacitor bank once over the horizon, each with its own labor hours and component mass. Cost is given by Eq. \eqref{eq:maint_cost}.

\begin{equation}
C_{\mathrm{maint}} = \sum_{k=1}^{K}\left[ \sum_{y=0}^{T-1} \frac{h_{\mathrm{pm}} r + c_{\mathrm{cons}}}{(1+r_d)^{y}}
+ \sum_{j}\sum_{y \in \mathcal{Y}_j} \frac{h_j r + m_j c_j}{(1+r_d)^{y}} \right]
\label{eq:maint_cost}
\end{equation}
where $r = \$95$/h is a fully burdened maintenance rate, $\mathcal{Y}_j$ is the set of replacement years for component $j$, and $m_j$ and $c_j$ are its mass and unit cost. Maintenance carbon is the embodied carbon of the replacement parts at their material factors, plus a small per hour allowance for technician travel and shop overhead which is carried in the sensitivity rather than in the central claim. The rate, the hours, the intervals and the labor carbon allowance are each swept low, central and high.

\subsection{Reliability Model}
\label{sec:reliability_model}

We assigned each subloop count to an uptime institute tier downtime target. That mapping described the redundancy a topology is built to provide, not the failure rate of its trains, and the two were not distinguished. The model is rebuilt here from component data so that the aggregation is explicit.

Each subloop train is a series arrangement of pump, VFD and isolation valve. With mean times between failure from IEEE Std 493 \cite{ieee493} of 50,000, 100,000 and 200,000 h respectively and a uniform 8 h mean time to repair (MTTR) reflecting the swap and repair practice at ORNL, the per component unavailability is $q_j = \mathrm{MTTR}/(\mathrm{MTBF}_j + \mathrm{MTTR})$ and the train unavailability is calculated by Eq. \eqref{eq:train_unavail}.

\begin{equation}
q_{\mathrm{train}} = 1 - \prod_{j} (1 - q_j) = 2.80\times10^{-4}
\label{eq:train_unavail}
\end{equation}

It equivalent to 2.45 h/yr. Aggregation to the plant depends on the redundancy policy. Without a standby, loss of any duty train degrades the plant.

\begin{equation}
q_{\mathrm{plant}}^{\,\mathrm{(no\ standby)}} = 1 - (1 - q_{\mathrm{train}})^{K}
\label{eq:plant_series}
\end{equation}
where one N+1 standby able to take over within the repair time, the cooling function is lost only when a second train fails during that window, giving the two out of $K+1$ term.

\begin{equation}
q_{\mathrm{plant}}^{\,\mathrm{(N+1)}} = 1 - (1-q)^{n} - n\,q\,(1-q)^{n-1}, \quad n = K+1
\label{eq:plant_n1}
\end{equation}

Equations \eqref{eq:plant_series} and \eqref{eq:plant_n1} replace the tier assignment used previously. The tier targets are retained only as an external reference scale for stating a policy requirement, not as a computed quantity.

Downtime is not energy neutral, and the coupling is modeled rather than neglected. With one duty train out of service the remaining $K-1$ carry the same heat load, so the plant hydraulic resistance rises and pump power at fixed total flow rises by $(K/(K-1))^{2}$. The extra electricity over the horizon is
\begin{equation}
\Delta E = \bar{P}_{\mathrm{pump}} \left[ \left(\tfrac{K}{K-1}\right)^{2} - 1 \right] D_{\mathrm{yr}} T
\label{eq:downtime_energy}
\end{equation}
where $\bar P_{\mathrm{pump}}$ is mean pumping power and $D_{\mathrm{yr}}$ is annual degraded running hours. This term is carried through to the life cycle totals in Section \ref{sec:reliability_results}.

The model covers per subloop failure modes only. Common cause events at the CEP-B level, such as electrical service interruption or a chemistry excursion, affect all $K$ approximately equally and would shift every candidate by a similar amount without changing the ranking. Recent reliability modeling of liquid cooling systems for data center availability supports treating the pump and drive as the dominant contributors at this level of the plant \cite{jayatilleka2026}.
}

\subsection{Life Cycle Aggregation and Multi-objective Formulation}
\label{sec:multi_obj}

\revised{For each candidate partition the three objectives:
\begin{align}
G(\mathbf{n}) &= G_{\mathrm{emb}} + G_{\mathrm{maint}} + \textstyle\sum_{y=0}^{T-1} E_{\mathrm{op}}(\mathbf{n})\,\mathrm{CI}(y)\label{eq:obj_carbon}\\
C(\mathbf{n}) &= C_{\mathrm{cap}} + C_{\mathrm{maint}} + \textstyle\sum_{y=0}^{T-1} \frac{E_{\mathrm{op}}(\mathbf{n})\,p_e}{(1+r_d)^{y}} \label{eq:obj_cost}\\
D(\mathbf{n}) &= 8{,}760\,q_{\mathrm{plant}}(K) \label{eq:obj_downtime}
\end{align}
where $E_{\mathrm{op}}$ is annual operational electricity, $\mathrm{CI}(y)$ is the grid carbon intensity in year $y$, $p_e$ is the electricity price and $r_d$ the discount rate. Capital cost enters at full value in year zero because it is a single expenditure at construction.}

Equations \eqref{eq:obj_carbon} to \eqref{eq:obj_downtime} define the three objectives. A partition $\mathbf{n}_a$ dominates $\mathbf{n}_b$ if $C(\mathbf{n}_a) \leq C(\mathbf{n}_b)$, $G(\mathbf{n}_a) \leq G(\mathbf{n}_b)$ and $D(\mathbf{n}_a) \leq D(\mathbf{n}_b)$ with at least one strict. The Pareto set is the subset that nothing dominates.

\subsection{Semi-analytical Decision Rule and the Redundancy Policy}
\label{sec:decision_rule}

To generalize beyond Frontier, the life cycle objective is written analytically in terms of two system parameters, the total HTW peak flow $Q$ and the drop budget $N_{\mathrm{CDU}} \times L_{\mathrm{drop}}$, with coefficients calibrated to the bill of materials and the operational model. \revised{Reliability enters this rule as a feasibility filter on $K$ and as a cost term, not as a weighted objective, following the treatment of reliability as a system level design constraint \cite{mcdonald2008}. Three policies are defined:}

\revised{
\begin{itemize}
\item \textbf{P0}, no redundancy requirement. Every $K$ is feasible and the objective contains duty trains only.
\item \textbf{P1}, N+1 standby required. One standby train is added to every candidate. Its embodied carbon and capital cost enter the objective and depend on $K$, because the standby is sized to take over the largest duty branch and that branch shrinks as $K$ grows.
\item \textbf{P2}, N+1 standby and hydraulic ride through. As P1, and additionally the surviving $K-1$ duty branches must carry full plant flow within 1.35 times design velocity, which imposes a lower bound on $K$.
\end{itemize}
}

The rule is evaluated on a $40 \times 40$ grid spanning $Q \in [50, 1{,}000]$ kg/s and $NL \in [25, 600]$ m. \revised{Because the pipe schedule is quantized, the objective is not monotone in $K$ and several values can lie within a fraction of a percent of the best, so the near-optimal set within 1\% is reported alongside the minimizer rather than hidden by a single argument.}

\subsection{Sensitivity Analysis Design}
\label{sec:sensitivity_design}

A one at a time sweep covers 5 dimensions at 3 levels each, giving 15 scenarios: operational horizon at 5, 7 and 10 year; TVA grid trajectory optimistic, central and pessimistic; discount rate at 3, 5 and 7\%; material cost at minus 20\%, central and plus 20\%; and embodied carbon factor at minus 30\%, central and plus 30\%. The material cost bounds reflect the historical spread of US steel and copper prices over 5 year and the embodied bounds reflect the published uncertainty in the ICE database supplemented by the spread across available EPDs. \revised{A two parameter sweep varies the embodied factor and grid intensity together, and Sections \ref{sec:layout}, \ref{sec:scalability} and \ref{sec:datareq} add three further robustness tests on physical layout, on the search method, and on the operational data requirement.}
\section{Results}
\label{sec:results}

\subsection{Reference Configuration and Headline Result}
\label{sec:headline}

\revised{The as built reference is stated first, because the saving depends on it. The public dataset description records that CEP-B contains four cooling subloops of which numbers 1 to 3 are in service and number 4 is a backup \cite{sun2024frontier}, and the companion co-design study reports the in-service partition as $(14,6,5)$ \cite{Jadhav2026CoDesign}. The reference used here is three duty subloops carrying 14, 6 and 5 CDUs, with a fourth standby train. Previously, we used a balanced four subloop proxy $(7,6,6,6)$; that proxy is not the built plant and the comparison has been recomputed.}

The cost and carbon optimum is $K=2$ with partition $(14,11)$, at 3,320.7 t CO\textsubscript{2}e and \$3,987k over the 7 year horizon. \revised{Against the as built duty configuration at 3,356.3 t and \$4,050k this is a saving of 35.7 t CO\textsubscript{2}e (1.06\%) and \$63k (1.56\%). Including maintenance the saving becomes 36.0 t and \$80k, the cost difference widening because each retired train removes about \$16.5k of maintenance net present value.}

\begin{table}[h]
\caption{Best partition per subloop count $K$, ranked by total life cycle CO\textsubscript{2}e. \revised{Maintenance is included. The as built duty configuration is shown for reference on the second to last row; the saving is computed from unrounded totals.}}
\label{tab:per_k_best}
\centering
\begin{tabular}{ccccc}
\toprule
$K$ & Partition & CO\textsubscript{2}e (t) & Cost (k\$) & \revised{Maint. (k\$)} \\
\midrule
2 & (14, 11)            & 3,321.3 & 4,020 & 33.1 \\
3 & (11, 8, 6)          & 3,336.9 & 4,074 & 49.6 \\
4 & (8, 8, 7, 2)        & 3,341.8 & 4,111 & 66.1 \\
5 & (11, 8, 2, 2, 2)    & 3,355.3 & 4,164 & 82.6 \\
6 & (8, 8, 3, 2, 2, 2)  & 3,369.5 & 4,219 & 99.2 \\
\midrule
\revised{3} & \revised{(14, 6, 5) built} & \revised{3,357.3} & \revised{4,099} & \revised{49.6} \\
\multicolumn{2}{l}{\revised{Saving over built}} & \revised{36.0} & \revised{80} & \revised{16.5} \\
\bottomrule
\end{tabular}
\end{table}

Table \ref{tab:per_k_best} shows the best partition at each $K$. The penalty for moving away from $K=2$ grows monotonically but stays small in relative terms, from 15.6 t and \$54k at $K=3$ to 48.2 t and \$199k at $K=6$.

\subsection{Contributions by Life Cycle}
\label{sec:contributions}

\begin{figure}[t]
\centering
\includegraphics[width=0.8\linewidth]{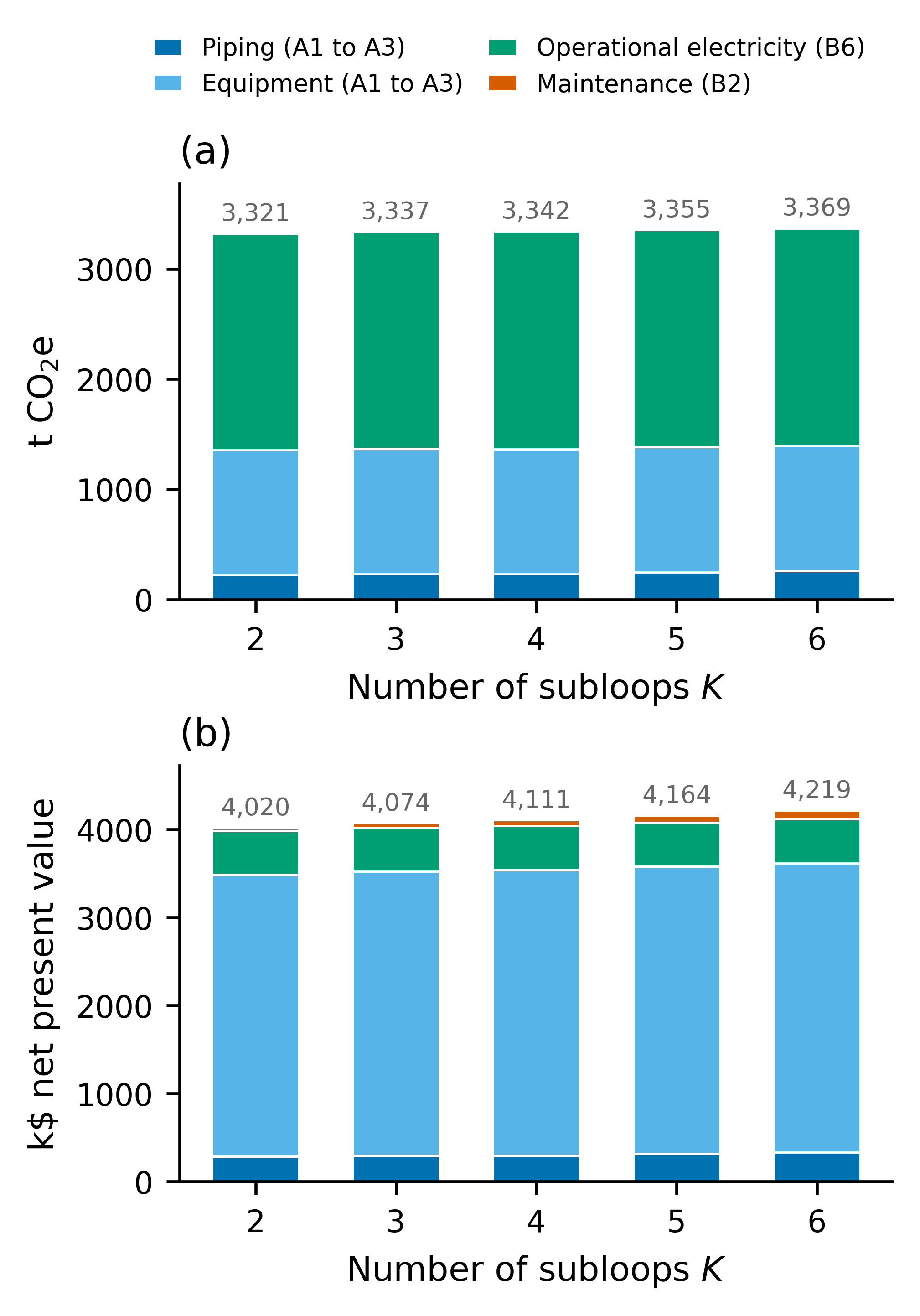}
\caption{\revised{Life cycle carbon and cost by stage for the best partition at each $K$. (a) Life cycle carbon: operational electricity dominates at about 1,970 t and equipment is constant at 1,134 t, so the variation with $K$ comes almost entirely from piping; maintenance is included but stays under 2 t and is not visible at this scale. (b) Life cycle cost: equipment dominates, and the variation with $K$ comes from piping, equipment sizing and maintenance together.}}
\label{fig:stagebreak}
\end{figure}

\revised{Figure \ref{fig:stagebreak} decomposes both objectives by stage.} Operational electricity contributes about 1,970 t across all $K$, equipment a constant 1,134 t, and piping rises from 222.6 t at $K=2$ to 261.0 t at $K=6$. \revised{On cost the picture inverts: equipment dominates at about \$3,200k to \$3,284k, operational electricity contributes about \$500k, piping \$286k to \$335k, and maintenance \$33k to \$99k.} Operational carbon is essentially insensitive to $K$: the median difference between $K=2$ and $K=6$ is about 9.4 MWh/yr, which is \revised{0.8\% of annual operational energy}. The embodied difference that separates partitions comes from piping, because splitting the system flow into more loops replicates trunk and header runs and standard pipe schedules quantize the resulting bore.

\revised{Maintenance behaves differently from the other terms and is worth separating. It scales with the number of maintainable trains, so it falls monotonically as $K$ falls: 310 technician hours and \$33.1k of net present value at $K=2$, against 930 hours and \$99.2k at $K=6$. Reducing $K$ from 6 to 2 saves \$66.1k of maintenance and 1.23 t CO\textsubscript{2}e of replacement parts. The cost effect is material, comparable to the entire piping difference across the same range, while the carbon effect is negligible against a 3,300 t total. Maintenance reinforces the preference for fewer duty trains on cost and does not move it on carbon.}

\revised{This insensitivity is produced by the per timestep optimizer rather than assumed, which is what justifies its cost. Strategies B and C differ only in whether the flow fractions are held proportional to CDU count or solved at each timestep. Moving from B to C reduces the median $K=2$ to $K=6$ gap from 12.17 to 9.36 MWh/yr and the median within-$K$ spread from 20.51 to 15.80 MWh/yr, both by 23\%. It has a larger effect on the quantity that matters for the design claim, the gap in annual energy between the as built plant and the best partition: 12.72 MWh/yr under proportional flow against 3.27 MWh/yr under per timestep optimization, a closure of 74\%. Establishing this cost about 7 CPU-min per partition and 71 CPU-h in total. Section \ref{sec:datareq} shows that the partition choice itself does not require this fidelity, so the expense buys the finding rather than the design.}

\revised{One consequence deserves note because it separates this study from its operational predecessor. The partition that minimizes annual cooling energy is $(19,6)$ at 1,173.68 MWh/yr, marginally ahead of $(14,11)$ at 1,173.79 MWh/yr, and it is the optimum reported in the companion co-design study \cite{Jadhav2026CoDesign}. On life cycle terms $(19,6)$ is not competitive, at 3,392.3 t against 3,320.7 t, because placing 19 CDUs on one branch forces a larger bore over the full routed length and adds 71.8 t of embodied carbon to save 0.11 MWh/yr. The operationally optimal topology and the life cycle optimal topology are different objects, and the difference is embodied.}

\revised{
\subsection{Physical CDU Assignment and Routing}
\label{sec:layout}

\begin{figure}[t]
\centering
\includegraphics[width=0.8\linewidth]{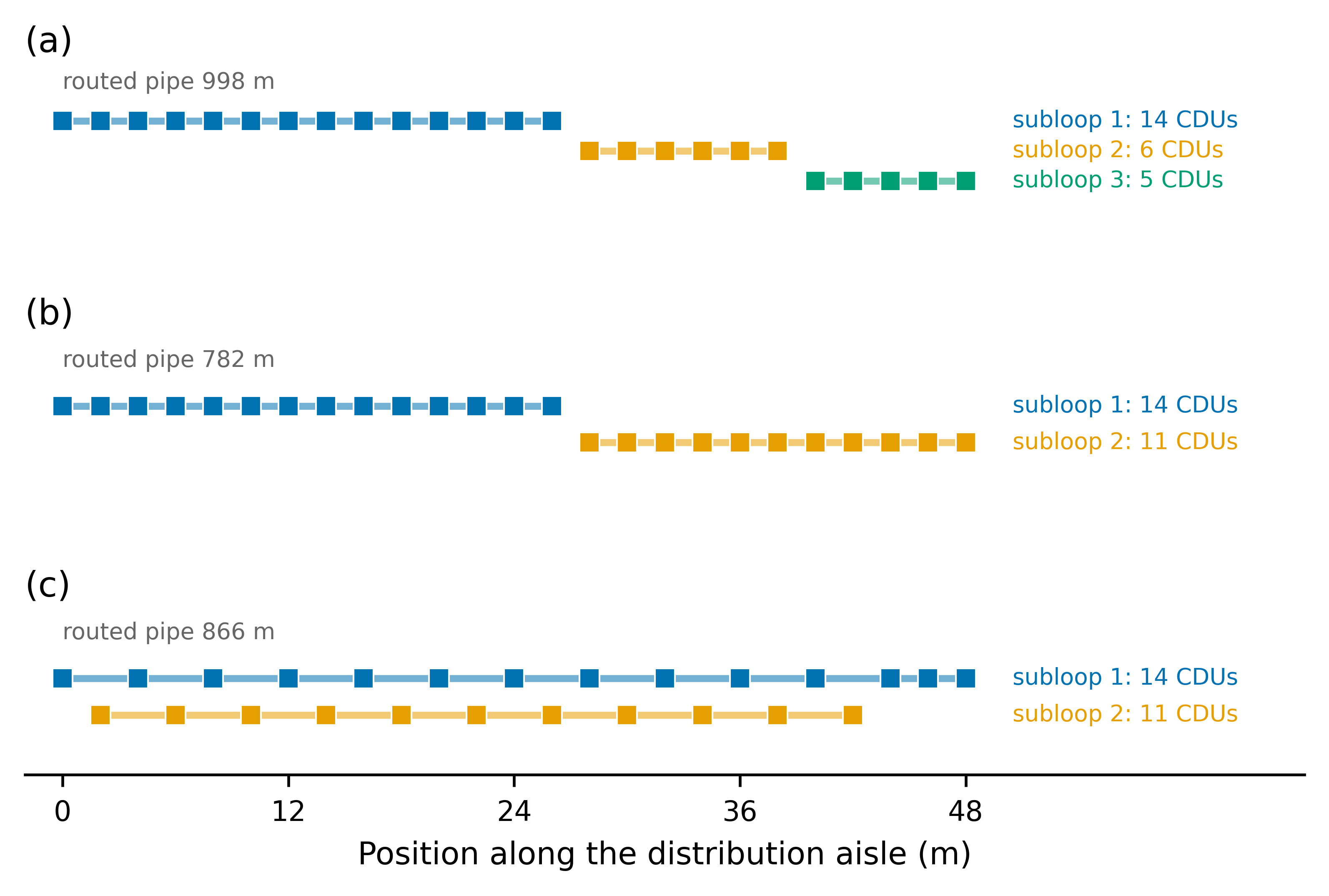}
\caption{\revised{Physical CDU assignment for a count based partition. The 25 CDUs lie along a 48 m distribution aisle. (a) The as built three duty subloops $(14,6,5)$ under contiguous grouping. (b) The life cycle optimum $(14,11)$ under contiguous grouping. (c) The same partition $(14,11)$ under interleaved grouping. A partition fixes the counts, not the grouping; (b) and (c) are the two extremes of grouping for one partition, and the routed pipe length differs by 84 m between them.}}
\label{fig:assignment}
\end{figure}

A partition specifies counts, not which physical CDUs are grouped, so each entry in Table \ref{tab:per_k_best} stands for a family of layouts. That family is bracketed here with an explicit floor plan model. The 25 CDUs are placed along a distribution aisle at 2.0 m spacing, and the header length of a subloop is the span of its assigned CDUs plus a 5 m tie back to the trunk at each end. Two extremes bound the family: contiguous grouping, where each subloop takes one unbroken block of the aisle, and interleaved grouping, where subloops are assigned round robin so each spans the whole aisle. Figure \ref{fig:assignment} shows both for the optimum.

Routed pipe length under contiguous grouping runs from 782 m at $K=2$ to 1,646 m at $K=6$; under interleaved grouping it runs from 866 m to 2,026 m. The fixed 50 m header allowance of Eq. \eqref{eq:route_len} gives 850 m to 2,050 m, so the base model sits close to the interleaved extreme. The published pipe figures are the pessimistic end of the routing family, not the optimistic end.

The consequence for the design conclusion is worth stating plainly. Under interleaved grouping the best partition at each $K$ gives 3,325, 3,335, 3,339, 3,342 and 3,354 t for $K=2$ to 6, a 28 t spread. Under contiguous grouping the same sequence is 3,303, 3,306, 3,305, 3,315 and 3,320 t, a 17 t spread in which $K=2$, 3 and 4 lie within 3 t of one another. The identity of the optimum is unchanged at $(14,11)$ under both extremes, and the saving over the as built configuration is 22.8 t under contiguous routing and 26.2 t under interleaved routing. The ordering is robust; the margin is not large, and a compact physical grouping removes most of the penalty for building more subloops.

Hydraulic loss was computed with Darcy-Weisbach on the selected bore using the Swamee-Jain friction factor. The worst branch loss is 5.15 m under contiguous grouping and 5.70 m under interleaved grouping at $K=2$, against the 120 m total dynamic head assumed for the pumps. A scattered layout thus costs about 2.6 kW of additional pumping at peak flow. Routing affects the embodied term more than the operational one.

\subsection{Redundancy Policy and the As Built Configuration}
\label{sec:reliability_results}

\begin{figure}[t]
\centering
\includegraphics[width=0.8\linewidth]{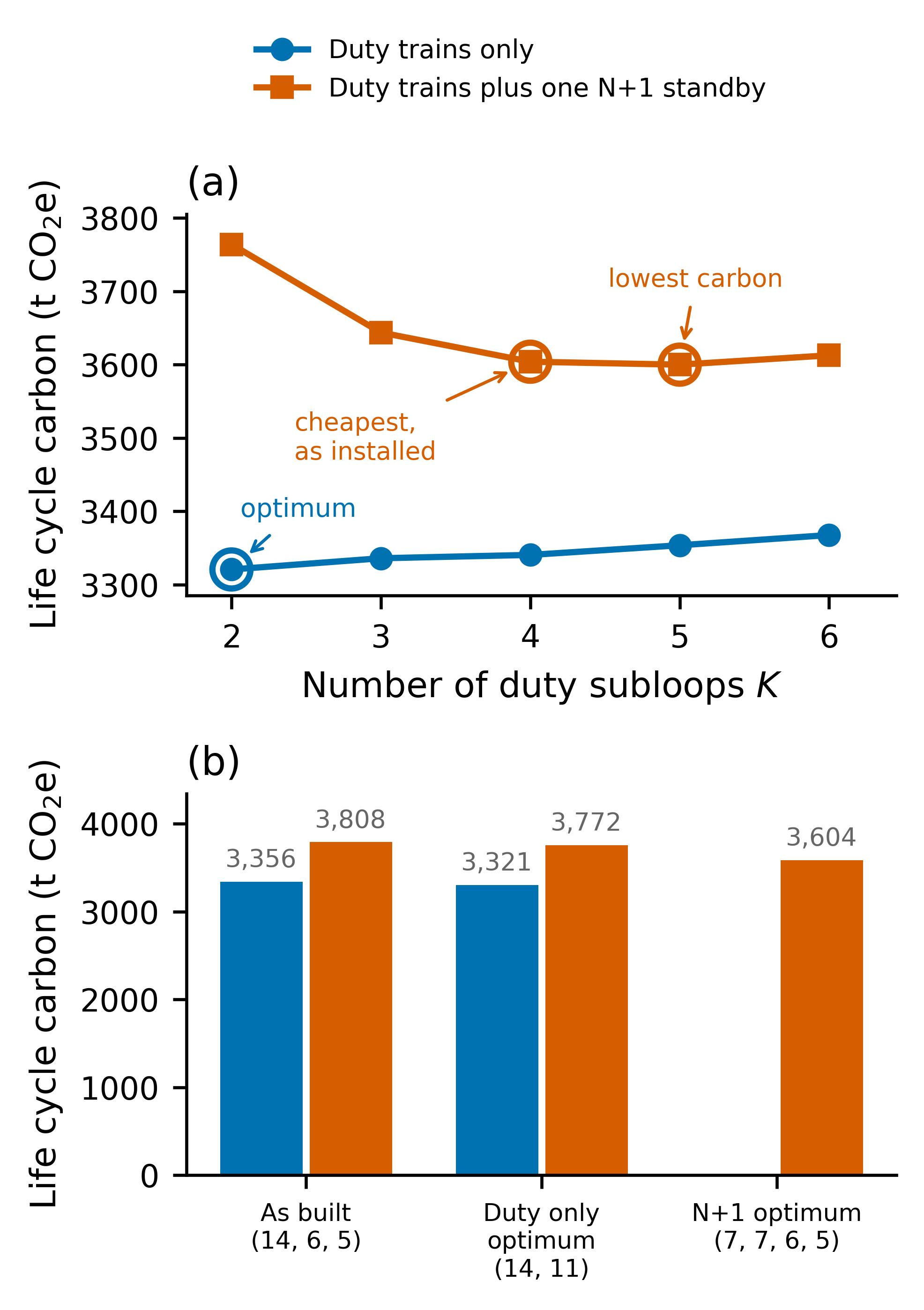}
\caption{\revised{Effect of pricing the standby train on life cycle carbon. (a) Best partition at each duty subloop count, with and without one N+1 standby. Without a standby the minimum is unambiguous at $K=2$. With a standby the curve is flat between $K=4$ and $K=5$: $K=4$ is the cheaper of the two at \$4,424k against \$4,433k, and $K=5$ is lower on carbon by 4.0 t, so both are marked. (b) The as built configuration, the duty only optimum, and the optimum under an N+1 requirement. Costing the redundancy moves the optimum from two duty subloops to four or five; the as built configuration stays above all of them because its largest branch forces an oversized spare.}}
\label{fig:redundancy}
\end{figure}

The results above price the duty trains only. Frontier carries a fourth train as a standby, and pricing it changes the answer.

A standby must be able to take over any single duty train, so it is sized to the largest duty branch. That branch shrinks as $K$ grows, so the standby is expensive at low $K$ and cheap at high $K$. Applying one N+1 standby uniformly to all 611 partitions gives best-per-$K$ carbon totals of 3,764.1 t at $K=2$ with $(13,12)$, 3,643.8 t at $K=3$ with $(9,8,8)$, 3,604.0 t at $K=4$ with $(7,7,6,5)$, 3,600.0 t at $K=5$ with $(7,7,7,2,2)$ and 3,612.7 t at $K=6$ with $(6,6,6,3,2,2)$. \revised{The as built duty configuration with its standby sits at 3,807.6 t, above every one of these, because its largest branch of 14 CDUs forces a standby train of 451.3 t and \$622k.} The optimum moves from $K=2$ to $K \in \{4,5\}$, with $K=6$ also inside the 1\% near-optimal band, and $K=2$ becomes the worst option rather than the best, as shown in Fig. \ref{fig:redundancy}.

The result is not an artifact of the standby sizing convention. Sizing the standby to an equal share of plant flow, as a common header arrangement of $K+1$ identical pumps would give, or to the mean duty branch, moves the optimum to $K=6$ at 3,530.9 t. Under all three conventions every $K \ge 4$ dominates $K=2$.

The reliability aggregation of Eqs. \eqref{eq:train_unavail} to \eqref{eq:plant_n1} explains the reason. A single train has 2.45 h/yr of expected downtime. Without a standby, plant downtime grows with the duty count, from 4.90 h/yr at $K=2$ to 14.70 h/yr at $K=6$, because more trains in series means more ways to lose one. With one N+1 standby it collapses to between 0.002 and 0.014 h/yr across the same range. Availability is bought by the spare, not by subdividing the duty.

The energy coupling of Eq. \eqref{eq:downtime_energy} is small but not zero. At the modeled exposure the extra pumping electricity over the horizon is 890 kg CO\textsubscript{2}e at $K=2$ and 396 kg at $K=6$, so degraded running costs low $K$ about 0.5 t more over the horizon, roughly 1\% of the 35.7 t design difference. It would take about 70 h/yr of degraded running before this term reached 10\% of the design difference. The effect is reported for completeness and does not change any ranking.

\revised{These three results separate two questions that the unconstrained analysis had merged. On duty count the built plant sits close to the constrained recommendation: three duty trains plus a standby against the four or five the framework selects, and far from the two that cost and carbon alone give. On CDU assignment it does not. The built partition $(14,6,5)$ with its standby reaches 3,807.6 t, which is 163.8 t above the best three subloop partition $(9,8,8)$ at 3,643.8 t and 203.6 t above the constrained optimum, because concentrating 14 CDUs on one branch forces a spare sized to match it. What the redundancy requirement recovers is the plant's duty count, not its assignment, and the cost and carbon advantage of two duty subloops exists only for a facility willing to run without a spare.}
}

\subsection{Sensitivity and Robustness}
\label{sec:sensitivity}

\begin{figure}[!t]
\centering
\includegraphics[width=0.8\linewidth]{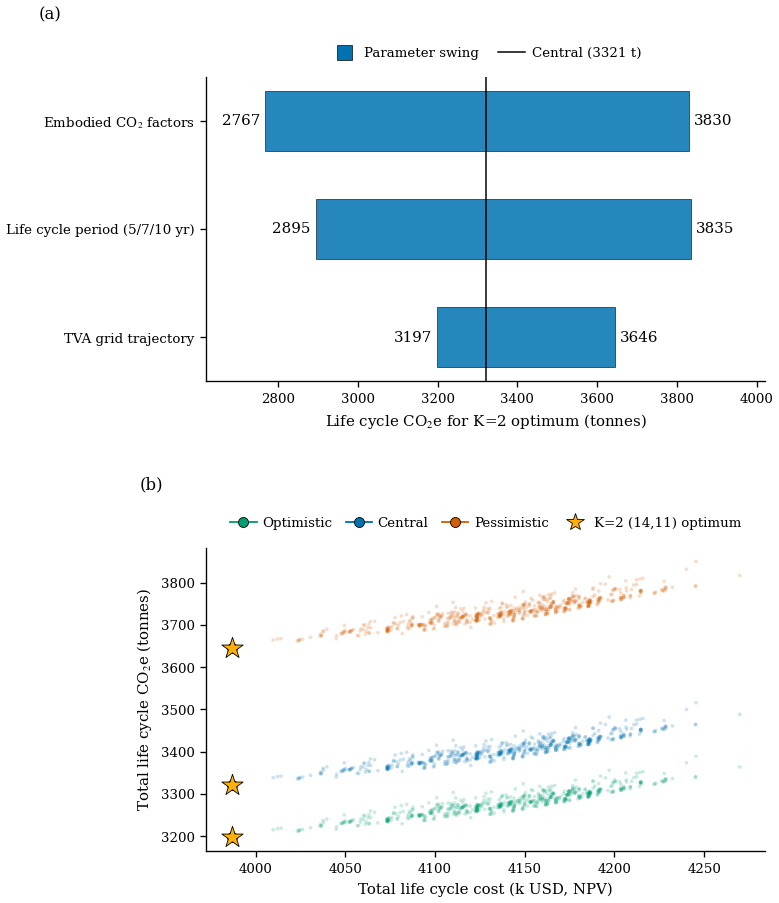}
\caption{Sensitivity and robustness of the $K=2$ optimum. (a) Tornado decomposition of life cycle CO\textsubscript{2}e for the $(14,11)$ partition across \revised{the three sensitivity dimensions that move the carbon balance; the discount rate and material cost dimensions move only the cost objective and are omitted from the tornado}, with the vertical line at the central value of 3,321 t. (b) Pareto front under three TVA grid trajectory scenarios; the gold star marks the optimum in each.}
\label{fig:sensitivity}
\end{figure}

\begin{figure}[t]
\centering
\includegraphics[width=0.8\linewidth]{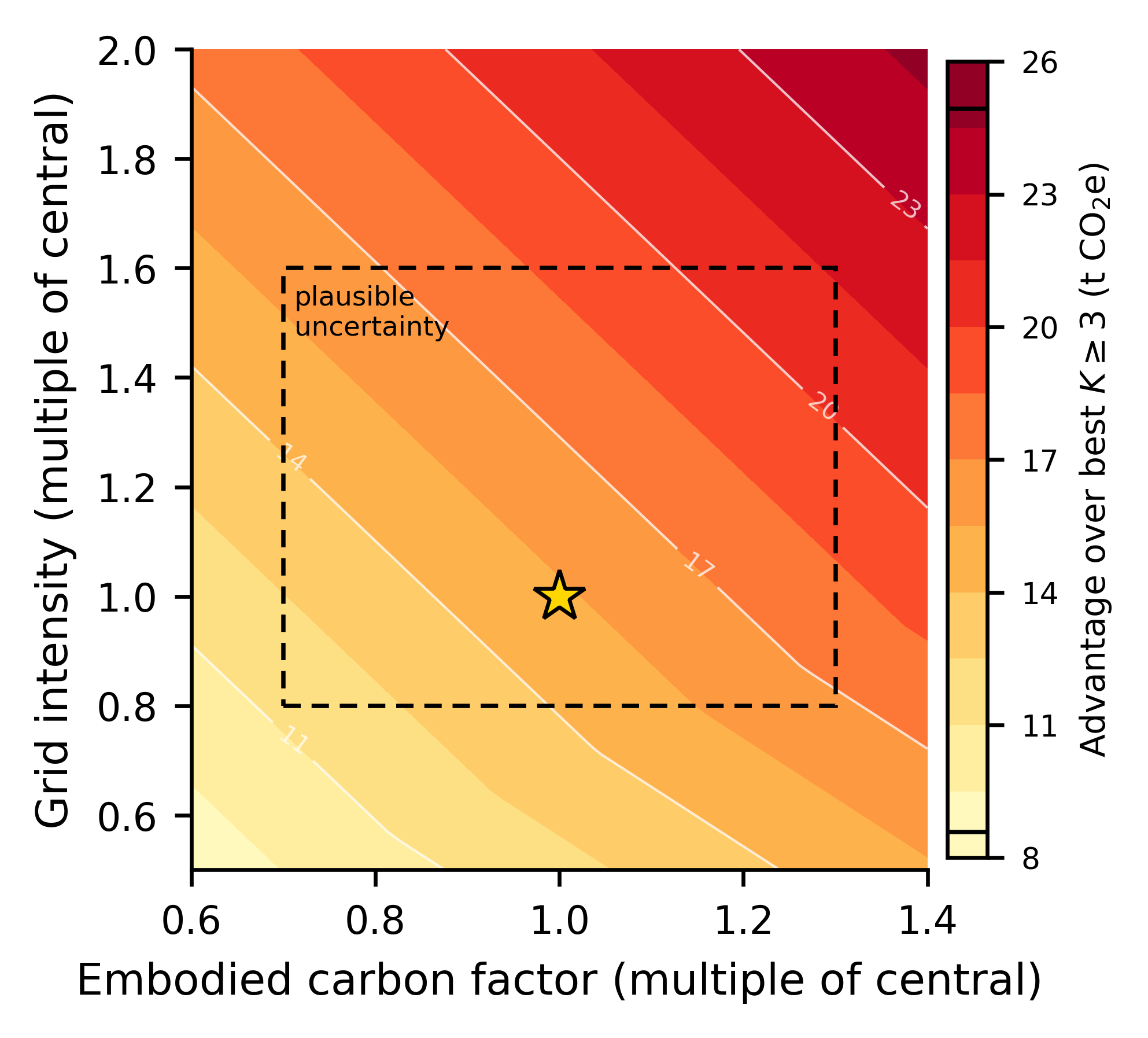}
\caption{\revised{Two parameter sensitivity. Advantage of the $K=2$ partition $(14,11)$ over the best $K\geq3$ partition in life cycle CO\textsubscript{2}e as a function of embodied carbon factor and grid intensity, each as a multiple of its central value. The dashed box marks the plausible uncertainty range and the star the central case, where the advantage is 15.3 t. The advantage stays positive across the whole sweep, so $(14,11)$ remains the optimum throughout.}}
\label{fig:twoparam}
\end{figure}

Figure \ref{fig:sensitivity}(a) decomposes the life cycle carbon of the $(14,11)$ optimum across \revised{the three sensitivity dimensions that move it}. Embodied carbon factors are the largest contributor, with the life cycle total ranging from 2,767 t to 3,830 t around the central 3,321 t. The horizon is second (2,895 to 3,835 t) and the grid trajectory third (3,197 to 3,646 t). Discount rate and material cost do not enter the carbon balance. In all 15 scenarios the $(14,11)$ partition remains on the Pareto front, and Fig. \ref{fig:sensitivity}(b) shows it holding that position under all three grid trajectories. Figure \ref{fig:twoparam} extends this to a two parameter test: across a $\pm40\%$ range on embodied factor and a factor of two range on grid intensity, $(14,11)$ remains the optimum with an advantage over the best $K\geq3$ partition ranging from \revised{8.6 to 24.9 t}.

\revised{These sensitivities test parameter uncertainty at fixed structure. The structural tests are elsewhere: routing in Section \ref{sec:layout}, redundancy policy in Section \ref{sec:reliability_results} and operational model fidelity in Section \ref{sec:datareq}. The redundancy policy is the only one of the three that changes the answer, which is the central finding of this revision.}

\subsection{Further Analyses of the $K=2$ Optimum}
\label{sec:deepdives}

\begin{figure}[t]
\centering
\includegraphics[width=\textwidth]{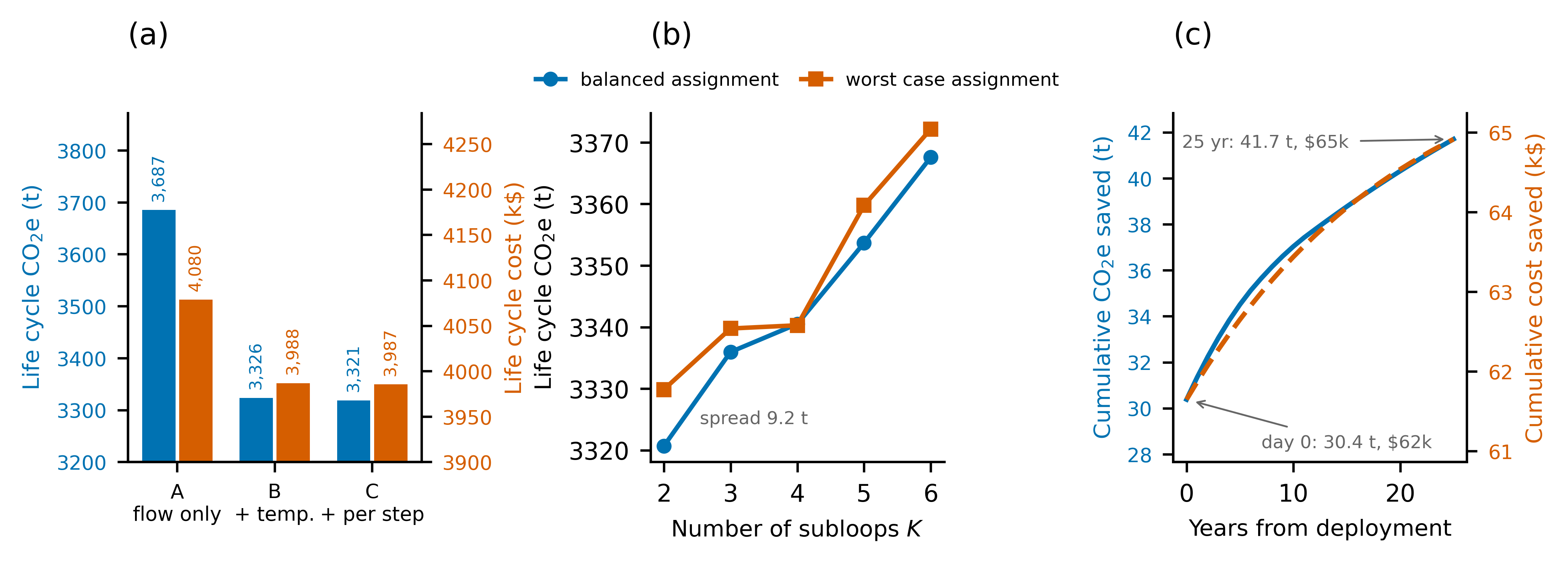}
\caption{\revised{Three analyses on the $K=2$ optimum. (a) Life cycle CO\textsubscript{2}e and cost under Strategies A, B and C. (b) Life cycle CO\textsubscript{2}e at each $K$ for the best partition under the balanced and the adversarial CDU assignment. (c) Cumulative saving of $(14,11)$ over the documented as built configuration $(14,6,5)$ across 25 year. Panel (c) has been redrawn against the corrected reference.}}
\label{fig:tier1}
\end{figure}

\revised{Figure \ref{fig:tier1} addresses three further questions.} Life cycle carbon falls from 3,687 t under Strategy A to 3,326 t under Strategy B to 3,321 t under Strategy C, and cost from \$4,080k to \$3,988k to \$3,987k. The choice of optimum is unaffected: $(14,11)$ is lowest under every strategy. \revised{The balanced and adversarial CDU assignments differ by 9.2 t for the best $K=2$ partition. Across all 611 partitions the adversarial penalty is bounded above by 15.3 t, reached at $K=6$; at some partitions it is slightly negative, because the adversarial assignment maximizes thermal imbalance and the per timestep flow optimizer can occasionally exploit that imbalance. In every case the effect is small against the 3,321 t total and does not reorder the partitions.}

\revised{Panel (c) shows that the saving is realized almost entirely at construction. The embodied and capital difference of 30.4 t and \$61.6k is present on day zero, and 25 year of operation adds only a further 11.3 t and \$3.3k, because the operational difference between the two topologies is small under Strategy C. This is the same asymmetry that makes a retrofit uneconomic in Section \ref{sec:retrofit}.}

\subsection{Decision Map and Generalization}
\label{sec:decisionmap}

\begin{figure}[t]
\centering
\includegraphics[width=\textwidth]{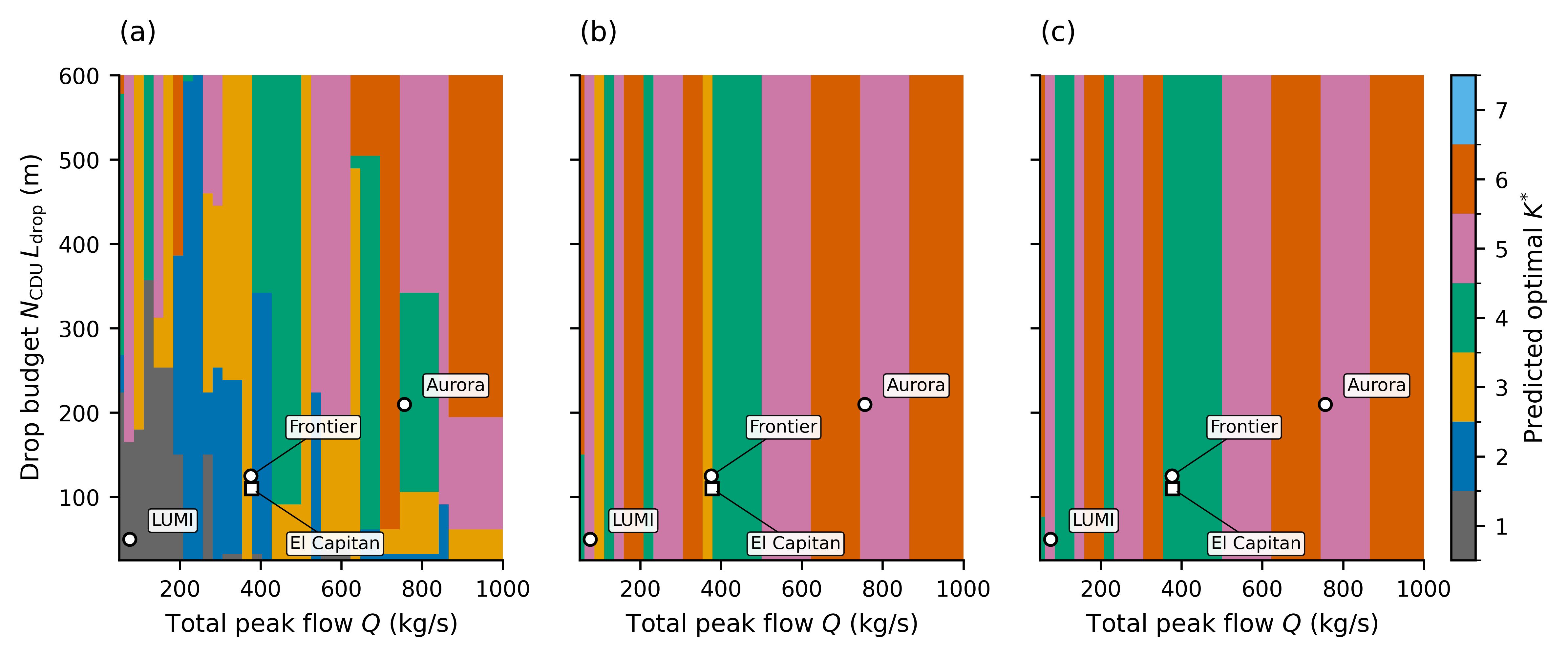}
\caption{\revised{Predicted optimal $K^{\ast}$ against total HTW peak flow and drop budget, under three availability policies: (a) no redundancy requirement, (b) N+1 standby required, and (c) N+1 standby together with hydraulic ride through. Markers show the four systems of Table \ref{tab:systems}; El Capitan is drawn as a square because it lies within 15 kg/s and 15 m of Frontier. Costing the N+1 standby moves the prediction for Frontier from two duty subloops to four, adjacent to the three the plant runs. The maps differ on 50.2\% of grid cells between (a) and (b) and on 5.3\% between (b) and (c).}}
\label{fig:decisionmap}
\end{figure}

\revised{Figure \ref{fig:decisionmap} shows the decision map responds to the reliability requirement. Under P0 the predicted optima are $K^{\ast}=2$ for Frontier and El Capitan, 4 for Aurora and 1 for LUMI. Under P1 they become 4, 5, 4 and 6, with the near-optimal sets within 1\% being $\{4,5\}$, $\{5,6\}$, $\{4,5\}$ and $\{6\}$. Adding the hydraulic ride through requirement of P2 changes only 5.3\% of the map, because at the flow rates of interest the velocity constraint is slack wherever the N+1 requirement has already pushed $K$ up. The regime boundaries move substantially: under P0 the map is spread across $K$ from 1 to 6, with no single value holding more than 21.4\% of the cells; under P1 no cell selects $K \le 2$ and 77\% of cells select $K \ge 5$.}

\revised{For Frontier specifically the analytical objective under P1 falls from \$4,580k and 3,809 t at $K=2$ to \$4,397k and 3,629 t at $K=4$, then flattens. Under the unconstrained policy the rule pointed at two duty subloops, well below what the plant runs; under the N+1 policy it points at four, adjacent to it. The constrained rule is consistent with the built plant in a way the unconstrained rule was not.}

\subsection{Three Objective Pareto Behavior}
\label{sec:pareto3d}

\begin{figure}[!t]
\centering
\includegraphics[width=0.8\linewidth,height=0.72\textheight,keepaspectratio]{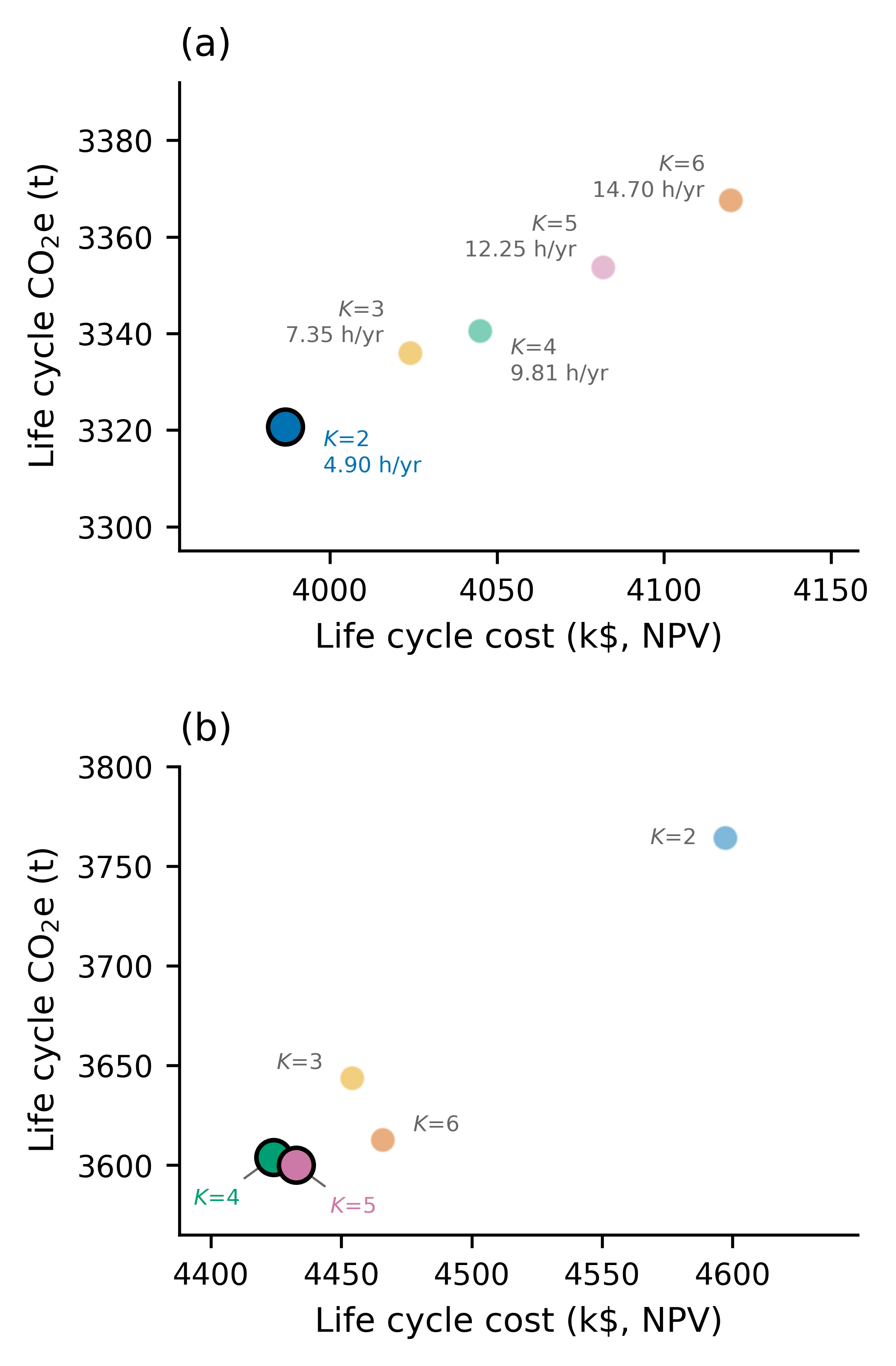}
\caption{\revised{Three objective structure for Frontier under the two redundancy policies, using the component based downtime of Section \ref{sec:reliability_model}. Each marker is the best partition at that duty subloop count, and non-dominated configurations are outlined. (a) Without a standby, $K=2$ leads on cost, carbon and downtime simultaneously and the Pareto set is a single point; the annotation below each marker is its expected unplanned downtime, which grows from 4.90 h/yr at $K=2$ to 14.70 h/yr at $K=6$. (b) With one N+1 standby, expected downtime falls below 0.015 h/yr for every candidate, from 0.002 h/yr at $K=2$ to 0.014 h/yr at $K=6$. Those differences are seconds per year, far below the resolution of the IEEE Std 493 component data, so they are not treated as discriminating and are omitted from the markers; the remaining cost against carbon front contains $K=4$, which is cheaper, and $K=5$, which is lower on carbon.}}
\label{fig:pareto3d}
\end{figure}

\revised{With the component based downtime of Eqs. \eqref{eq:plant_series} and \eqref{eq:plant_n1}, the three objective structure separates cleanly by policy, and Fig. \ref{fig:pareto3d} shows both cases for Frontier, where the full 611 partition evaluation exists. Without a standby, low $K$ is best on all three objectives simultaneously, since it is cheapest, lowest carbon and has the fewest duty trains to fail; the Pareto set collapses to the single point $K=2$ and there is no trade to make. With an N+1 standby, downtime falls below 0.015 h/yr for every $K$. The residual downtime differences between candidates amount to seconds per year and sit far below the uncertainty of the component failure data, so they are not treated as discriminating, and the front on the two remaining objectives contains $K=4$, which is cheaper, and $K=5$, which is lower on carbon. The genuine trade is not between subloop count and availability but between providing a spare and not providing one; once the spare is provided, the choice among $K \ge 4$ is a small cost against carbon trade. The cross system generalization is carried by the decision maps of Fig. \ref{fig:decisionmap}.}

\subsection{Search Cost and Scalability}
\label{sec:scalability}

Enumeration is exact at $N=25$ but will not remain practical as plants grow. The number of partitions of $N$ into 2 to 6 parts is 611 at $N=25$, 9,191 at $N=50$, 189,508 at $N=100$, 4,775,382 at $N=200$ and 134,851,968 at $N=400$. The count itself is polynomial for bounded $K$; the binding cost is the operational evaluation attached to each candidate, which for this study was about 7 CPU-min per partition over 49,353 timesteps, 71 CPU-h in total. The life cycle layer on top of it runs in 0.02 ms per partition.

Three alternatives were implemented and their optimality gap measured against full enumeration on the $N=25$ problem, where the exact answer is known, as shown in Table \ref{tab:search}.

\begin{table}[h]
\caption{\revised{Alternatives to full enumeration, measured against the exact optimum on the $N=25$ problem.}}
\label{tab:search}
\centering
\small
\begin{tabular}{lccc}
\toprule
Method & Evaluations & Selected & Carbon gap \\
\midrule
Full enumeration & 611 & (14, 11) & exact \\
Bounded imbalance $\delta \le 3$ & 26 & (14, 11) & 0.000\% \\
Bounded imbalance $\delta \le 1$ & 5 & (13, 12) & 0.551\% \\
Greedy from balanced & 209 & (14, 11) & 0.000\% \\
Decision rule then balanced & 2 & (13, 12) & 0.551\% \\
\bottomrule
\end{tabular}
\end{table}

Restricting the search to partitions whose largest and smallest branches differ by at most three CDUs reduces the candidate set from 611 to 26 and still finds the exact optimum. This restriction is defensible on physical grounds independently of the result, since a strongly unbalanced partition oversizes one branch and wastes the capacity of the others. Greedy refinement from the balanced partition at each $K$, moving one CDU at a time while the objective improves, also recovers the exact optimum using 34\% of the full evaluations. The decision rule alone requires two evaluations and lands 0.55\% away.

The bounded imbalance and greedy candidate sets do not grow with $N$ in the way full enumeration does: the bounded set has 26 members for any $N$ at $K \le 6$, and greedy requires at most $K(K-1)$ moves per pass. A plant with several hundred CDUs would be searched by bounded enumeration or greedy refinement, with the rule used for the initial $K$ screen. Where each evaluation is expensive enough that even these are impractical, surrogate assisted search of the kind developed for expensive engineering design \cite{bhattacharjee2018} is the natural next step, and heuristic methods have been used at comparable scale for district cooling network connection \cite{neri2022district}.

\subsection{Operational Data Needed for Decision}
\label{sec:datareq}

The framework used a full year of measurements from an operating machine. A designer of a next generation system has no such record, so the question is what part of it the decision actually requires.
The design decision enters through differences in the operational term between candidates, and those differences are small: annual energy spans 1,173.7 to 1,205.3 MWh across the 611 partitions, a range of 2.69\%. Two error models were tested. Under independent per partition error, which assumes the model error between two partitions differing by a single CDU move is uncorrelated, the exact optimum survives 0.5\% error 61\% of the time and 1\% error 19\% of the time. That is the worst case and it is not the realistic one, because a designer applies one model to every candidate and the resulting error is shared. Splitting the error into a common and a partition specific component and sweeping the correlation, at correlation 0.99 and 2\% absolute error the exact optimum is recovered in 98\% of trials with zero median regret, and at 5\% absolute error in 66\% of trials, again with zero median regret.

The systematic test confirms this. Replacing the Strategy C operational term with Strategy B, with Strategy A, with a constant equal to the median, or with the benchmark baseline of 1,819 MWh/yr returns partition $(14,11)$ in every case, with Spearman rank correlation against the full ranking of 0.97 in the worst case. Substituting a design load estimate computed from nameplate duty and a design point pump curve, with no operating record at all, also returns $(14,11)$.
What the decision requires is consistency across candidates rather than accuracy in absolute terms. The operational record is needed to state absolute energy and carbon totals and to establish that operational energy is nearly independent of topology; it is not needed to select the partition. This makes the framework applicable to a next generation design, and Section \ref{sec:retrofit} explains the difficulties of applying to an existing one.

\subsection{Retrofit Possibility}
\label{sec:retrofit}

Converting the built plant from three duty subloops to two requires isolating, draining, flushing and passivating one branch, cutting out and removing its trunk and header pipework, re-tying its five CDU drops onto the surviving headers, relocating or abandoning one pump, drive and heat exchanger, and rebalancing and re-commissioning the plant. At 295 h of removal and tie-in labor, 18,093 kg of pipe removed, new tie-in material, and the associated engineering, permitting and contingency, the conversion costs about \$264k and emits 8.65 t CO\textsubscript{2}e under the cut-off convention.

The saving it can capture is much smaller than the design difference, because in a built plant the embodied and capital difference of 30.4 t and \revised{\$61.6k} is already committed. Only the operational difference remains, and that is 5.28 t and \$1.34k over 7 years. The conversion costs about 197 times what it saves in money, with a simple payback of about 1,378 year, and about twice what it saves in carbon over the horizon, with a carbon payback of 11.5 year. Even crediting the full greenfield difference rather than the operational part alone, the cost of conversion exceeds the benefit by a factor of 4.2. Under the avoided burden convention the retrofit carbon becomes negative at $-7.64$ t, and the allocation choice is stated in Section \ref{sec:framework_overview}; the cost verdict is unchanged either way.

This framework is a greenfield design tool. It can provide guidance for the next plant, more than a recommendation to modify a plant already built. The operational recommendations of the companion study \cite{Jadhav2026CoDesign}, which are software changes to flow allocation and supply temperature setpoints, remain the appropriate intervention for an existing facility.

\section{Discussion}
\label{sec:discussion}

\revised{The result that survives every test in this study is not that two subloops are optimal. It is that cost and carbon discriminate weakly between subloop counts, and that the redundancy policy discriminates strongly.}
On cost and carbon alone the optimum is $K=2$ with partition $(14,11)$, and the advantage over the built configuration is 35.7 t CO\textsubscript{2}e and \$63k over 7 years. \revised{That advantage narrows to about 2 t against the best four subloop design once a compact physical grouping of CDUs is assumed (Section \ref{sec:layout}), and it reverses entirely once the standby train the plant carries is priced (Section \ref{sec:reliability_results}).} Operational energy is nearly invariant to $K$ under Strategy C because the optimizer extracts essentially all available flexibility at $K=2$, so the marginal cost of an additional subloop is not recovered in operations. The differences that remain are embodied, and specifically piping.

\revised{The design guidance that follows is conditional on the redundancy policy and can be stated in one line: choose the smallest duty subloop count that meets the hydraulic and thermal requirement, provide N+1 redundancy explicitly rather than by subdividing the duty, and group the CDUs on each subloop contiguously and in near equal blocks. For a Frontier class plant carrying a spare this gives four or five duty trains, one or two more than were built, with the duty spread evenly rather than concentrated on a single branch of 14 units.}

Operators of new systems can read the predicted $K^{\ast}$ from Fig. \ref{fig:decisionmap} at their own peak flow and drop budget, \revised{using the panel that matches their availability policy. The policy matters more than the operating point: moving from no redundancy requirement to an N+1 requirement changes the prediction on half the map, whereas moving Frontier's operating point by $\pm 25\%$ does not change it at all.} Operators in other electricity markets should rerun the rule with their local grid trajectory; cleaner grids shift the optimum toward lower $K$ and dirtier grids toward higher $K$, but this effect is second order against the redundancy policy.

\revised{
Treating reliability as a hard constraint set by operations policy, rather than as a soft objective traded against cost and carbon, is the formulation that makes this problem well posed \cite{mcdonald2008}. The three objective view placed downtime on an axis and invited a trade. The component level model shows that the trade is illusory once a standby is provided, because downtime then falls below 0.015 h/yr for every candidate and stops discriminating. What the designer is really choosing is whether to provide a spare, and that choice is made by facility policy, not by an optimizer. Once made, it enters as a term in the objective through the standby train's own embodied carbon and cost, and that term is what selects $K$.
}

First, embodied carbon factors come from a combination of EPDs and the ICE database. The high and low variants bracket published uncertainty but do not account for systematic upward revision of all factors as carbon accounting matures \cite{kota2010}. \revised{Second, the routing model of Section \ref{sec:layout} is a linear aisle abstraction, not the Frontier floor plan, which is not public at the resolution required; it brackets the routing family rather than reproducing the built route.} Third, the analysis assumes the per CDU heat load distribution observed in 2023 is representative of the service life. \revised{Fourth, the reliability model covers per subloop failure modes and a single N+1 standby with perfect switchover within the repair time; common cause failures, planned maintenance outages and controls faults are excluded, so the absolute downtime figures are lower bounds and should be read as relative comparisons. Fifth, cooling parameters for Aurora, El Capitan and LUMI are inferred from public specifications rather than disclosed by their operators, so their predictions are illustrative.} Sixth, the decision map is calibrated to Frontier specific parameters and operators elsewhere need to recalibrate.

\section{Conclusions}
\label{sec:conclusions}

This paper framed the cooling subloop architecture of an exascale supercomputer as a life cycle co-design problem rather than a constraint inherited from convention. \revised{Four conclusions follow.}

\revised{First, on cost and carbon alone the optimum for the Frontier cooling plant is two duty subloops with partition $(14,11)$, at 3,320.7 t CO\textsubscript{2}e and \$3,987k over a 7 year horizon, saving 35.7 t and \$63k against the documented as built configuration of three duty subloops carrying $(14,6,5)$. The difference is an embodied carbon result driven by piping, not an operational efficiency result, and it survives all 15 sensitivity scenarios and both extremes of physical CDU grouping.}

\revised{Second, that ranking reverses when the redundancy the plant actually carries is priced. A standby train sized to take over the largest duty branch is expensive at low subloop counts and cheap at high ones, and including it moves the optimum to at least four duty subloops under every sizing convention tested. The recommended duty count moves from two, well below what Frontier runs, to four or five, just above it, so what sets the subloop count is the redundancy policy rather than cost or carbon. The built partition itself is not optimal at its own duty count, because concentrating 14 CDUs on one branch forces an oversized spare, and a component level reliability model shows the reason: availability is bought by the spare, not by subdividing the duty.}

\revised{Third, the framework does not need a year of operating data to reach its design conclusion. What it needs is one operational model applied consistently to every candidate; five different operational terms, including a constant and a data free design estimate, all select the same partition. It also does not need full enumeration: restricting the search to partitions with bounded imbalance reduces 611 candidates to 26 and still finds the exact optimum, which is what makes the method applicable to plants an order of magnitude larger.}

\revised{Fourth, the framework is a greenfield tool. Converting the built plant would cost about \$264k against a capturable operational saving of \$1.34k over 7 years, because the embodied and capital difference is already committed. The appropriate intervention for an existing facility is operational, not structural.}

Several research directions follow. The framework can be extended beyond high temperature water cooling to immersion, two phase and rear door architectures, so that the partition choice is optimized jointly with the cooling modality. \revised{The reliability layer can be deepened from the N+1 model used here to a full per partition fault tree with common cause modes and explicit switchover dynamics.} Integration with marginal grid carbon signals at the dispatch level would let operational decisions respond to real time grid composition. Finally, the life cycle assessment methodology developed here can be turned into a standardized reporting protocol for HPC facilities, supporting fair comparison of cooling plant carbon performance across institutions.

\revised{
\section*{Data and Code Availability}
The operational data underlying this study are public. The Frontier energy dataset is published as a data descriptor in Scientific Data \cite{sun2024frontier} and covers the full 2023 calendar year at 10 min resolution, from which 49,353 validated intervals are used here. The measured variables used by this framework are: HTW supply temperature; per subloop return temperature and volumetric flow for the three duty subloops; total HTW flow; per subloop and total waste heat load; supercomputer total power; and facility accessory power. Ambient wet bulb temperature enters through the calibrated cooling tower approach rather than directly. The reduced order plant model, its calibration and its validation against ASHRAE Guideline 14 are given in \cite{Jadhav2026Stage1}, and the per partition operational optimization in \cite{Jadhav2026CoDesign}. Material and cost factors are from the published ICE database \cite{ice_v41}, CIBSE TM65 \cite{cibse_tm65} and Turton \cite{turton2018}, all cited at the point of use. The life cycle, reliability, maintenance and decision rule code, together with the per partition operational results for all 611 partitions, will be made public available once the paper is accepted.
}

\bibliographystyle{unsrt}
\bibliography{References}

\end{document}